\pdfoutput=1
\documentclass[11pt,twoside,a4paper,cmspaper,final,collab]{cms-tdr}
\def\svnVersion{6548f07-D}\def\svnDate{2020/12/24}\def\cmsCernNoTag{CERN-EP-2020-141}\def\cmsCernDate{\today}\def\cmsMessage{"Published in Physics Letters B as \href{http://dx.doi.org/10.1016/j.physletb.2020.136036}{\doi{10.1016/j.physletb.2020.136036}.}"}
\begin{document}\cmsNoteHeader{HIN-19-009}

\newcommand {\roots}    {\ensuremath{\sqrt{s}}}
\newcommand {\deta}     {\ensuremath{\Delta\eta}}
\newcommand {\dphi}     {\ensuremath{\Delta\phi}}

\newcommand{\eecollision} {\ensuremath{\Pep\Pem}\xspace}
\newcommand{\epcollision} {\ensuremath{\Pe\Pp}\xspace}
\newcommand{\pp}     {\ensuremath{\Pp\Pp}\xspace}
\newcommand {\PbPb}  {\ensuremath{\text{PbPb}}\xspace}
\newcommand {\AuAu}  {\ensuremath{\text{AuAu}}\xspace}
\newcommand {\pPb}   {\ensuremath{\Pp\text{Pb}}\xspace}
\newcommand {\AonA}  {\ensuremath{\text{AA}}\xspace}
\newcommand {\pA}    {\ensuremath{\Pp\text{A}}\xspace}
\newcommand{\noff}    {\ensuremath{N_\text{trk}^\text{offline}}\xspace}
\newcommand{\Dzerodecay}{\ensuremath{\PDz\to\PKm\PGpp}\xspace}
\newcommand{\Dzerobardecay}{\ensuremath{\PADz\to \PKp\PGpm}\xspace}
\newcommand{\Dtopipi}{\ensuremath{\PDz\to \PGpp\PGpm}\xspace}
\newcommand{\DtoKK}{\ensuremath{\PDz\to \PKp\PKm}\xspace}

\newlength\cmsTabSkip\setlength{\cmsTabSkip}{1ex}
\newlength\cmsFigWidth
\ifthenelse{\boolean{cms@external}}{\setlength\cmsFigWidth{0.49\textwidth}}{\setlength\cmsFigWidth{0.9\linewidth}}

\cmsNoteHeader{HIN-19-009} 
\title{Studies of charm and beauty hadron long-range correlations in \pp and \pPb collisions at LHC energies}
\date{\today}

\abstract{
Measurements of the second Fourier harmonic coefficient ($v_2$) of the azimuthal distributions  of prompt and nonprompt \PDz mesons produced  in \pp and \pPb collisions are presented. Nonprompt \PDz mesons come from beauty hadron decays. The data samples are collected by the CMS experiment  at nucleon-nucleon center-of-mass energies of 13 and 8.16\TeV, respectively. In high multiplicity \pp collisions, $v_2$ signals for prompt charm hadrons are reported for the first time, and are found to be
comparable to those for light-flavor hadron species over a transverse momentum (\pt) range of 2--6\GeV. Compared at similar event multiplicities, the prompt \PDz meson $v_2$ values in \pp and \pPb collisions are similar in magnitude. The $v_2$ values for open beauty hadrons are extracted for the first time via nonprompt \PDz mesons in \pPb collisions.  For \pt in the range of 2--5\GeV, the results suggest that $v_2$ for nonprompt \PDz mesons is smaller than that for prompt \PDz mesons. These new measurements indicate a positive charm hadron $v_2$ in \pp collisions and suggest a mass dependence in $v_2$ between charm and beauty hadrons in the \pPb system. These results provide insights into the origin of heavy-flavor quark collectivity in small systems.}

\hypersetup{
pdfauthor={CMS Collaboration},
pdftitle={Studies of charm and beauty collectivity in pp and pPb collisions at LHC energies},
pdfsubject={CMS},
pdfkeywords={CMS, ridge, collectivity, small systems, heavy flavor, elliptic flow, charm, beauty}}

\maketitle
\section{Introduction}
\label{sec:intro}

Strong collectivity in high-energy nucleus-nucleus (\AonA) collisions
at the BNL RHIC~\cite{Ackermann:2000tr,Adler:2003kt,Ackermann:2000tr,Adams:2005ph,Alver:2008gk}
and at the CERN LHC~\cite{Aamodt:2010pa,Chatrchyan:2012ta},
has indicated the formation of a hot, strongly interacting quark gluon plasma (QGP), which exhibits 
nearly ideal hydrodynamic behavior~\cite{Ollitrault:1992bk,Heinz:2013th,Gale:2013da}. 
The collective phenomena manifests itself in long-range (large pseudorapidity gap) particle 
correlations~\cite{Alver:2009id,Abelev:2009af,Chatrchyan:2011eka,Chatrchyan:2012wg,ATLAS:2012at,CMS:2013bza}.
Although not originally expected, similar long-range collective azimuthal correlations
are also being observed in small colliding systems with high final-state particle multiplicity, such as proton-proton
(\pp)~\cite{Khachatryan:2010gv,Aad:2015gqa,Khachatryan:2015lva,Khachatryan:2016txc,Aad:2019aol}, 
proton-nucleus (\pA)~\cite{CMS:2012qk,alice:2012qe,Aad:2012gla,Aaij:2015qcq,ABELEV:2013wsa,Khachatryan:2014jra,Khachatryan:2015waa,
Aaboud:2017acw,Aaboud:2017blb,Aidala:2016vgl,PHENIX:2018lia},
and lighter nucleus-nucleus systems~\cite{Adamczyk:2015xjc,Adare:2015ctn,Aidala:2017ajz,PHENIX:2018lia}.
This observation raised the question of whether a fluid-like QGP medium 
with a size significantly smaller than in \AonA collisions is created in these other systems~\cite{Dusling:2015gta,Schlichting:2016sqo,Nagle:2018nvi}.
At the same time, there is no observation of long-range correlations in \eecollision and \epcollision collisions, 
which are even smaller systems compared to pp collisions~\cite{Badea:2019vey,ZEUS:2019jya}.
In the context of hydrodynamic models, the observed azimuthal correlation structure of emitted particles is
typically characterized by its Fourier components~\cite{Voloshin:1994mz}.
The second and third Fourier anisotropy coefficients are known as elliptic ($v_2$) and triangular ($v_3$) flow,
which most directly reflect the QGP medium response to the initial collision geometry and
its fluctuations, respectively~\cite{Alver:2010dn,Schenke:2010rr,Qiu:2011hf,Alver:2010gr}. 
The experimental measurements in the small systems are consistent with the dominance of
strong final-state interactions~\cite{He:2015hfa,Dusling:2015gta,Nagle:2018nvi,Bierlich:2018xfw,Kurkela:2019kip},
such as a hydrodynamic expansion of a tiny QGP droplet~\cite{Dusling:2015gta,Nagle:2018nvi}.
Alternative scenarios based on gluon saturation in the initial state can also capture the main 
features of the correlation data, and are conjectured to play a dominant role 
as the event multiplicity decreases~\cite{Dusling:2015gta,Schlichting:2016sqo}. 

Heavy-flavor quarks (charm and bottom) are produced via hard scatterings 
in the very early stages of the high energy collisions. These quarks are
available to probe both initial- and final-state effects of the collision dynamics~\cite{Andronic:2015wma,Dong:2019byy}. 
Strong elliptic flow signals of electrons from the decay of heavy-flavor hadrons and open charm \PDz mesons
are observed in both gold-gold (\AuAu) collisions at RHIC~\cite{Adare:2006nq,Adamczyk:2017xur}
and lead-lead (\PbPb) collisions at the LHC~\cite{Abelev:2014ipa,Acharya:2017qps,Sirunyan:2017plt}.
These findings suggest that charm quarks develop significant collective behavior via their strong interactions
with the bulk of the QGP medium. Measurements of elliptic flow of hidden-charm \PJGy
mesons provide further evidence for strong rescatterings of charm quarks~\cite{Khachatryan:2016ypw,Acharya:2017tgv}.

In small colliding systems, the study of heavy-flavor hadron collectivity has the potential
to disentangle possible contributions from both initial- and final-state 
effects. In particular, heavy flavor hadrons may be more sensitive to possible initial-state
gluon saturation effects. Recent observation of a significant elliptic flow signal for prompt 
\PDz~\cite{Sirunyan:2018toe} and prompt \PJGy~\cite{Acharya:2017tfn,Sirunyan:2018kiz} 
mesons in \pPb collisions provided the first evidence for charm quark collectivity 
in small systems. Surprisingly, despite the mass differences, the observed $v_2$ signal for prompt \PJGy mesons in \pPb collisions 
is found to be comparable to that of prompt \PDz mesons and light-flavor hadrons at a 
given particle transverse momentum (\pt). This behavior cannot be explained by the final-state effects
of a QGP medium, as the contribution from recombinations to \PJGy production is not expected to be significant in small systems~\cite{Du:2018wsj}.
This finding may imply the existence of initial-state correlation effects~\cite{Zhang:2019dth}.
Further detailed investigations are important to address many open questions for understanding
the origin of heavy-flavor quark collectivity in small systems. 
These include the multiplicity dependence of charm quark collectivity
in both \pPb and \pp systems and the details of collective behavior of beauty quarks.

This Letter presents the first measurement of the elliptic flow ($v_2$) for prompt \PDz
mesons in \pp collisions at center-of-mass energy $\roots = 13\TeV$ and for nonprompt \PDz mesons 
(from decays of beauty hadrons) 
in \pPb collisions at nucleon-nucleon center-of-mass energy $\sqrtsNN = 8.16\TeV$, using
long-range ($\abs{\deta}>1$) two-particle angular correlations. The $v_2$ harmonic coefficient is determined over
the 2--8\GeV \pt range for prompt \PDz mesons as a function
of multiplicity with results for the \pp and \pPb collisions.
The nonprompt \PDz meson $v_2$ values are
extracted in high-multiplicity \pPb collisions for two transverse momentum ranges
2--5 and 5--8\GeV,
and are compared to previous measurements of prompt \PDz mesons and light flavor hadrons.

\section{Experimental apparatus and data sample}
\label{sec:detector}

The central feature of the CMS apparatus is a superconducting solenoid of 6\unit{m}
internal diameter, providing a magnetic field of 3.8\unit{T}. Within the solenoid volume,
there are four primary subdetectors including a silicon pixel and strip tracker detector, 
a lead tungstate crystal electromagnetic calorimeter, and a brass and scintillator hadron 
calorimeter, each composed of a barrel and two endcap sections. Iron and quartz-fiber 
Cherenkov hadron forward calorimeters cover the pseudorapidity ($\eta_{\text{lab}}$) range 
 $2.9 < \abs{\eta_{\text{lab}}} < 5.2$ in laboratory frame.
Muons are measured in gas-ionization detectors embedded in the steel flux-return yoke 
outside the solenoid. The silicon tracker measures charged particles within the range 
$\abs{\eta_{\text{lab}}}< 2.5$. For charged particles with $1 < \pt < 10\GeV$ and $\abs{\eta_{\text{lab}}} < 1.4$, 
the track resolutions are typically 1.5\% in \pt and 25--90 (45--150)\mum in the transverse 
(longitudinal) impact parameter~\cite{Chatrchyan:2014fea}. A detailed description of the 
CMS detector, together with a definition of the coordinate system used and the relevant 
kinematic variables, can be found in Ref.~\cite{Chatrchyan:2008zzk}.

The event samples were collected by the CMS experiment with a two-level trigger system~\cite{Khachatryan:2016bia}:
at level-1 events are selected by custom hardware processors 
while the high-level trigger uses fast versions of the offline software.
The \pPb data at $\sqrtsNN  = 8.16\TeV$ used in this analysis were collected
in 2016, and correspond to an integrated luminosity of 186.0\nbinv~\cite{CMS-PAS-LUM-17-002}. The beam energies 
are 6.5\TeV for the protons and 2.56\TeV per nucleon for the lead nuclei. Because of the asymmetric beam conditions,
particles selected in this analysis from midrapidity in the laboratory frame ($\abs{y_{\text{lab}}}<1$) correspond
to rapidity in the nucleon-nucleon center-of-mass frame of $-1.46 < y_{\text{cm}} < 0.54$,
with positive rapidity corresponding to the proton beam direction. The  \pp data at $\roots  = 13\TeV$ 
were collected in 2017 and 2018 with integrated luminosities of 1.27\pbinv
and 10.22\pbinv during special runs with low beam intensity, 
resulting in an average number of 
concurrent  \pp collisions of about 1 per bunch crossing. The event reconstruction, event selections, and 
triggers (minimum bias and high multiplicity) 
are identical to those described in Refs.~\cite{Khachatryan:2016txc,Sirunyan:2017quh,Sirunyan:2017uyl}.
Similar to previous CMS correlation measurements, the \pPb and  \pp data are analyzed for several 
multiplicity (\noff) classes, where \noff is the number of offline selected tracks~\cite{Chatrchyan:2014fea,Khachatryan:2016txc} 
with $\abs{\eta_{\text{lab}}}<2.4$ and $\pt > 0.4\GeV$.

\section{Prompt and nonprompt \texorpdfstring{\PDz}{Dzero} meson reconstruction and selection}
\label{sec:data}

The \PDz (and its charge conjugate state \PADz) mesons are reconstructed through the hadronic decay channel {\Dzerodecay} ({\Dzerobardecay}).
The invariant mass of \PDz candidates is required to be from 1.725--2.000\GeV to cover the world-average \PDz mass~\cite{Zyla:2020zbs}.
In order to suppress the combinatorial background and improve the momentum and mass resolution,
high-purity~\cite{Chatrchyan:2014fea} tracks reconstructed using the
silicon tracker with $\pt > 0.7\GeV$, $\abs{\eta_{\text{lab}}} < 2.4$,
smaller than 10\% relative uncertainty in \pt, and the number of valid hits $\geq$11 are used. 
For each pair of selected tracks, two \PDz candidates are considered by assuming that
one of the tracks has the pion mass while the other track has the kaon mass, and vice versa.

The \PDz candidates are selected using a multivariate technique that employs the boosted decision 
tree (BDT) algorithm in the Toolkit for Multivariate Data Analysis with ROOT~\cite{Hocker:2007ht}.
The selection is optimized separately for \pp and \pPb collisions, and for all \pt ranges,
in order to maximize the statistical significance of the prompt 
or non-prompt \PDz meson signals. The Monte Carlo (MC) signal simulated samples are produced with
{\PYTHIA} 8.209~\cite{Sjostrand:2014zea} tune CUETP8M1~\cite{Khachatryan:2015pea}
(embedded into \textsc{epos lhc}~\cite{Pierog:2013ria} for the case of \pPb analysis) for both
prompt and nonprompt \PDz events. The background samples for the multivariate training are taken from data.
The training variables related to \PDz mesons include: the $\chi^2$ probability for \PDz vertex fitting;
the three-dimensional distance (with and without
being normalized by its uncertainty) between the primary and decay vertices; and the three-dimensional pointing angle
(defined as the angle between the line segment connecting the primary and decay vertices
and the momentum vector of the reconstructed particle candidates).
The training variables related to the decay products are: \pt; pseudorapidity and the longitudinal and transverse 
track impact parameter significance.
In the BDT training for prompt \PDz signals, same-sign (SS) \PGppm\PKpm candidates are used, which contain predominantly
combinatorial background. For optimizing nonprompt \PDz signals, both prompt \PDz signals and combinatorial
candidates are considered as dominant background to be suppressed. For this reason, opposite-sign (OS) candidates 
(although including fractions $<$5\% of nonprompt \PDz signals) are used for the background training sample.
This approach is found to give better performance for achieving higher nonprompt \PDz fractions 
than using SS background candidates, especially at higher \pt. 

The optimal selection criterion is the working point with the highest signal significance
of prompt and nonprompt \PDz signals. For extracting the nonprompt \PDz yield, the distributions 
of distance of closest approach (DCA) of the \PDz meson momentum vector, relative to the primary vertex, are fitted using the template 
probability distribution functions (PDs) for prompt and nonprompt \PDz signals derived from MC simulation. 
The residual nonprompt fraction in the BDT prompt-trained sample is found to be no more than 7\%, while in the 
BDT nonprompt-trained sample, the optimal selection yields a nonprompt fraction up to 20\%. 
This procedure is further outlined in Section ~\ref{sec:analysis}.

\section{Data analysis}
\label{sec:analysis}

The azimuthal anisotropies of \PDz mesons
are extracted from their long-range ($\abs{\deta}>1$) two-particle azimuthal 
correlations of \PDz candidates with charged particles, as described in Refs.~\cite{Khachatryan:2016txc,Khachatryan:2014jra}.
The two-dimensional (2D) correlation function
is constructed by pairing each \PDz candidate with reference primary charged-particle 
tracks with $0.3<\pt<3.0\GeV$ (denoted ``ref'' particles), and calculating
\begin{linenomath}
\begin{equation}
\frac{1}{N_{\PDz}}\frac{\rd^{2}N^\text{pair}}{\rd\deta\, \rd\dphi} = B(0,0)\ \frac{S(\deta,\dphi)}{B(\deta,\dphi)},
\end{equation}
\end{linenomath}
where $\deta$ and $\dphi$ are the differences in pseudorapidity $\eta_{\text{lab}}$ and azimuthal angle $\phi$ of each pair.
The same-event pair distribution, $S(\deta,\dphi)$, represents the yield of 
particle pairs normalized by the number of \PDz candidates ($N_{\PDz}$) from the same event. 
The mixed-event pair yield distribution, $B(\deta,\dphi)$,
is constructed by pairing \PDz candidates in each
event with the reference primary charged-particle tracks from 10 different randomly selected
events, from the same \noff\ range, and with a primary vertex falling in the same
2\cm wide range of reconstructed $z$ coordinates.
The $B(0,0)$ represents the value of $B(\deta,\dphi)$ at $\deta=0$ and $\dphi=0$.
It is evaluated by interpolating the four nearest bins with a bin width 
of $0.3$ in $\deta$ and $\pi/16$ in $\dphi$ bilinearly. 
The interpolation shows a negligible effect on the measurements.
The analysis procedure is performed in each \PDz candidate \pt range by dividing it into 14 intervals of invariant mass.
The correction for the acceptance and efficiency (derived from simulations using {\PYTHIA} for  \pp 
and {\textsc{pythia+epos}} for \pPb) of the \PDz meson yield
is found to have a negligible effect on the measurements, and is not applied. 
The corresponding effects are discussed in Section~\ref{sec:systematics}.
The \dphi\ correlation functions averaged over $\abs{\deta}>1$ (to
remove short-range correlations, such as jet fragmentation) are obtained from
the projection of 2D correlation functions and fitted by the first three terms of a Fourier series:
\begin{linenomath}
\begin{equation}
\label{eq:Vn}
\frac{1}{N_{\PDz}}\frac{\rd N^\text{pair}}{\rd\dphi} = \frac{N_{\text{assoc}}}{2\pi} \left[ 1+\sum_{n=1}^{3} 2V_{n\Delta} \cos (n\dphi)\right].
\end{equation}
\end{linenomath}
Here, $V_{n\Delta}$ are the Fourier coefficients and $N_{\text{assoc}}$
represents the total number of pairs per \PDz candidate.
The inclusion of additional Fourier terms to the fit has negligible effect.
By assuming $V_{n\Delta}$ to be the product of
single-particle anisotropies~\cite{Chatrchyan:2013nka}, 
$V_{n\Delta}(\PDz,{\text{ref}})=v_{n}(\PDz) v_{n}({\text{ref}})$,
the $v_n$ anisotropy harmonics for \PDz candidates can be extracted from the equation:
\begin{linenomath}
\begin{equation}
\label{eq:vn}
v_{n}(\PDz) = V_{n\Delta}(\PDz,{\text{ref}})\Big/\sqrt{V_{n\Delta}({\text{ref}},{\text{ref}})}.
\end{equation}
\end{linenomath}
\noindent Because of the limited statistical precision of the available data, only the elliptic anisotropy harmonic results are reported
in this analysis.

To extract the $V_{2\Delta}$ values of the inclusive \PDz meson signal ($V_{2\Delta}^{S}$),
a two-step fit to the invariant mass spectrum of \PDz candidates
and their $V_{2\Delta}$ as a function of the invariant mass, $V_{2\Delta}^{S+B}(m_{\text{inv}})$,
is performed in each \pt interval. The mass spectrum fit function is composed of
five components: the sum of two Gaussian functions with the same mean but different
widths for the \PDz signal, $S(m_{\text{inv}})$; an additional Gaussian function to describe the
invariant mass shape of \PDz candidates with an incorrect mass assignment from the
exchange of the pion and kaon designations, $SW(m_{\text{inv}})$; Crystal Ball (CB) functions~\cite{CrystalBallRef} to describe
processes $\Dtopipi$ ($S(m_{\PGpp\PGpm})$) and $\DtoKK$ ($S(m_{\PKp\PKm})$);
and a third-order polynomial
to model the combinatorial background, $B(m_{\text{inv}})$.
The contributions from the processes $\Dtopipi$ and $\DtoKK$
are the results of mislabelling \PK as \PGp, or vice versa. These two
components are emulated by two CB functions at two sides away from the peak region.
The width and the ratio of the yields of $SW(m_{\text{inv}})$ and
$S(m_{\text{inv}})$ and the CB function shape are fixed according to results obtained from simulation studies
using {\PYTHIA} for \pp collisions and {\textsc{pythia+epos}} for \pPb collisions.

The $V_{2\Delta}^{S+B}(m_{\text{inv}})$ distribution is fit with
\begin{linenomath}
\begin{equation}
V_{2\Delta}^{S+B}(m_{inv}) = \alpha(m_{inv})~V_{2\Delta}^{S} + [1-\alpha(m_\text{inv})]~V_{2\Delta}^{B}(m_\text{inv}),
\end{equation}
\end{linenomath}
where
\begin{linenomath}
\ifthenelse{\boolean{cms@external}}
{
\begin{multline}
   \alpha(m_{\text{inv}}) = \big [S(m_{\text{inv}}) + SW(m_{\text{inv}})+S(m_{\PKp\PKm}) \\
   +S(m_{\PGpp\PGpm}) \big ] \,/\, \big [ S(m_{\text{inv}}) + SW(m_{\text{inv}})\\
   +S(m_{\PKp\PKm})+S(m_{\PGpp\PGpm})+ B(m_{\text{inv}}) \big ]. 
\end{multline}
} 
{ 
\begin{equation}
\alpha(m_{\text{inv}}) = \frac{S(m_{\text{inv}}) + SW(m_{\text{inv}})+S(m_{\PKp\PKm})+S(m_{\PGpp\PGpm})}{
S(m_{\text{inv}}) + SW(m_{\text{inv}}) +S(m_{\PKp\PKm})+S(m_{\PGpp\PGpm})+ B(m_{\text{inv}})}.
\end{equation}
}
\end{linenomath}
Here $V_{2\Delta}^{B}(m_{\text{inv}})$ for the background \PDz candidates is modeled as a linear function
of the invariant mass, and $\alpha(m_{\text{inv}})$ is the \PDz signal fraction. The \PK-\Pgp
swapped, $\Dtopipi$ and $\DtoKK$ components are 
included in the signal fraction because these candidates are from genuine \PDz
mesons and should have the same $v_{2}$ value as that of the \PDz signal.

\begin{figure}[htb]
\centering
\includegraphics[width=\cmsFigWidth]{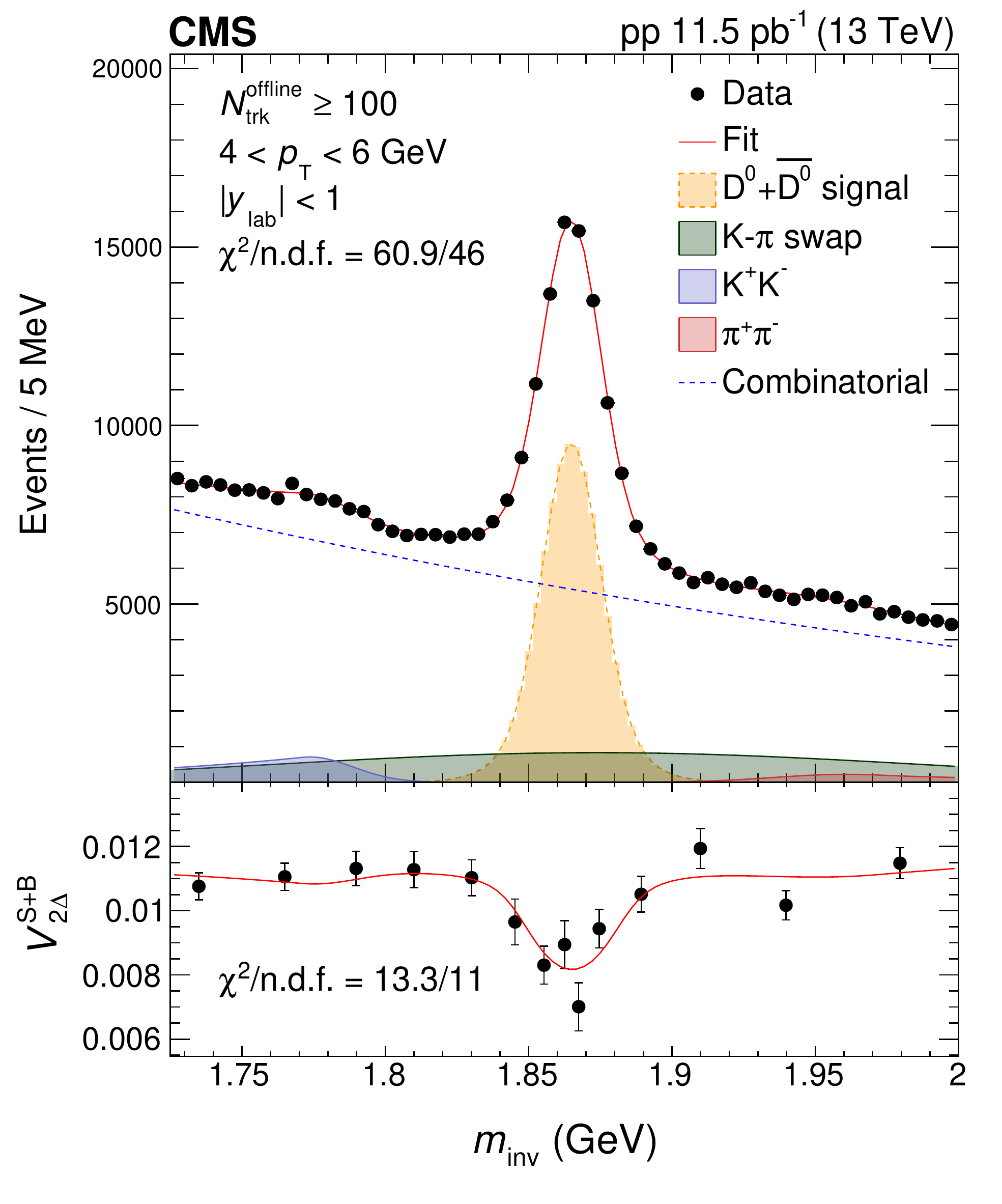}
\caption{Example of fits to the invariant mass spectrum and $V_{2\Delta}^{S+B}(m_{\text{inv}})$, 
for the BDT prompt-trained sample in  \pp collisions.}
\label{fig:massfitpp}
\end{figure}

\begin{figure*}[htb]
\centering
\includegraphics[width=0.48\textwidth]{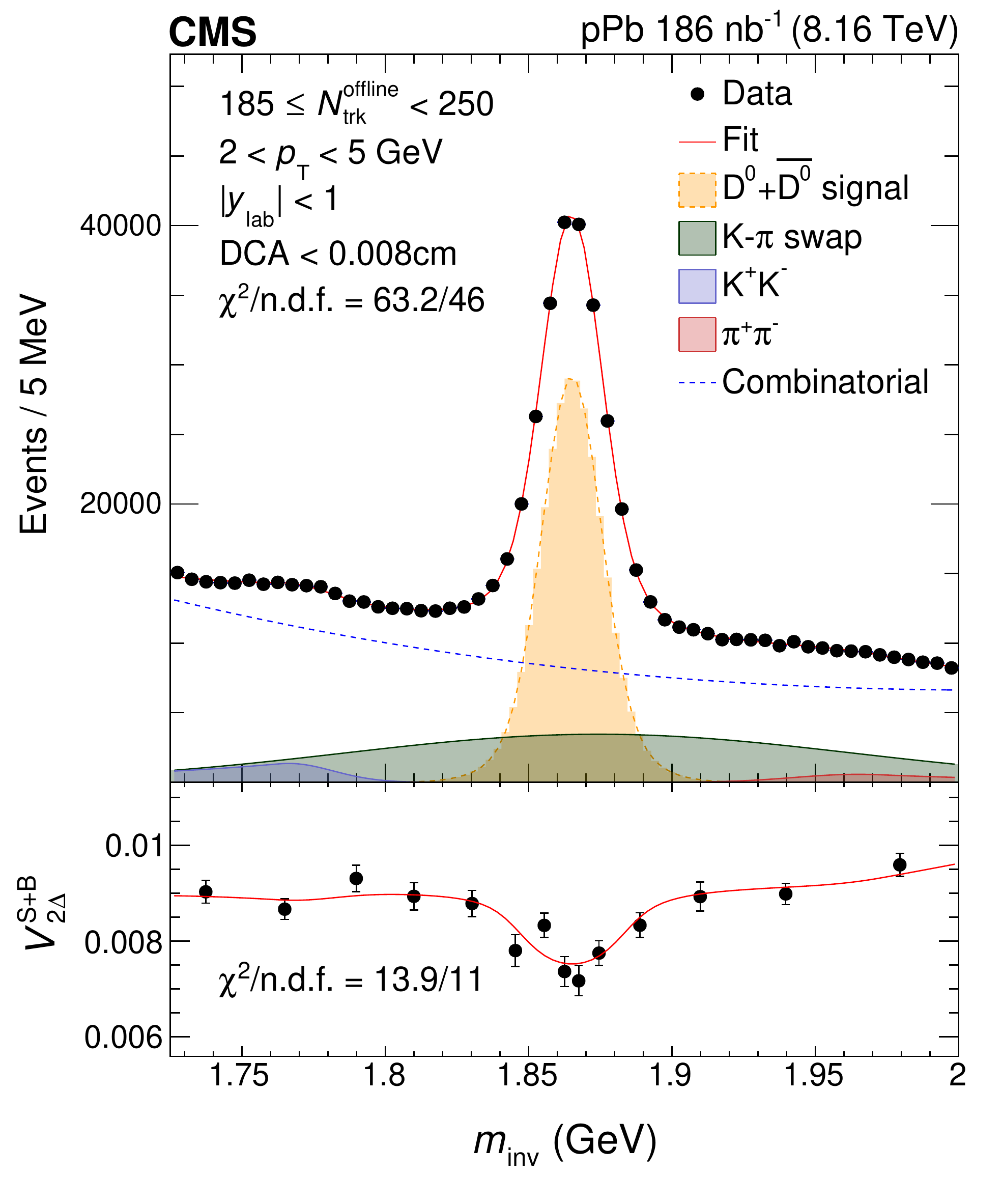}
\includegraphics[width=0.48\textwidth]{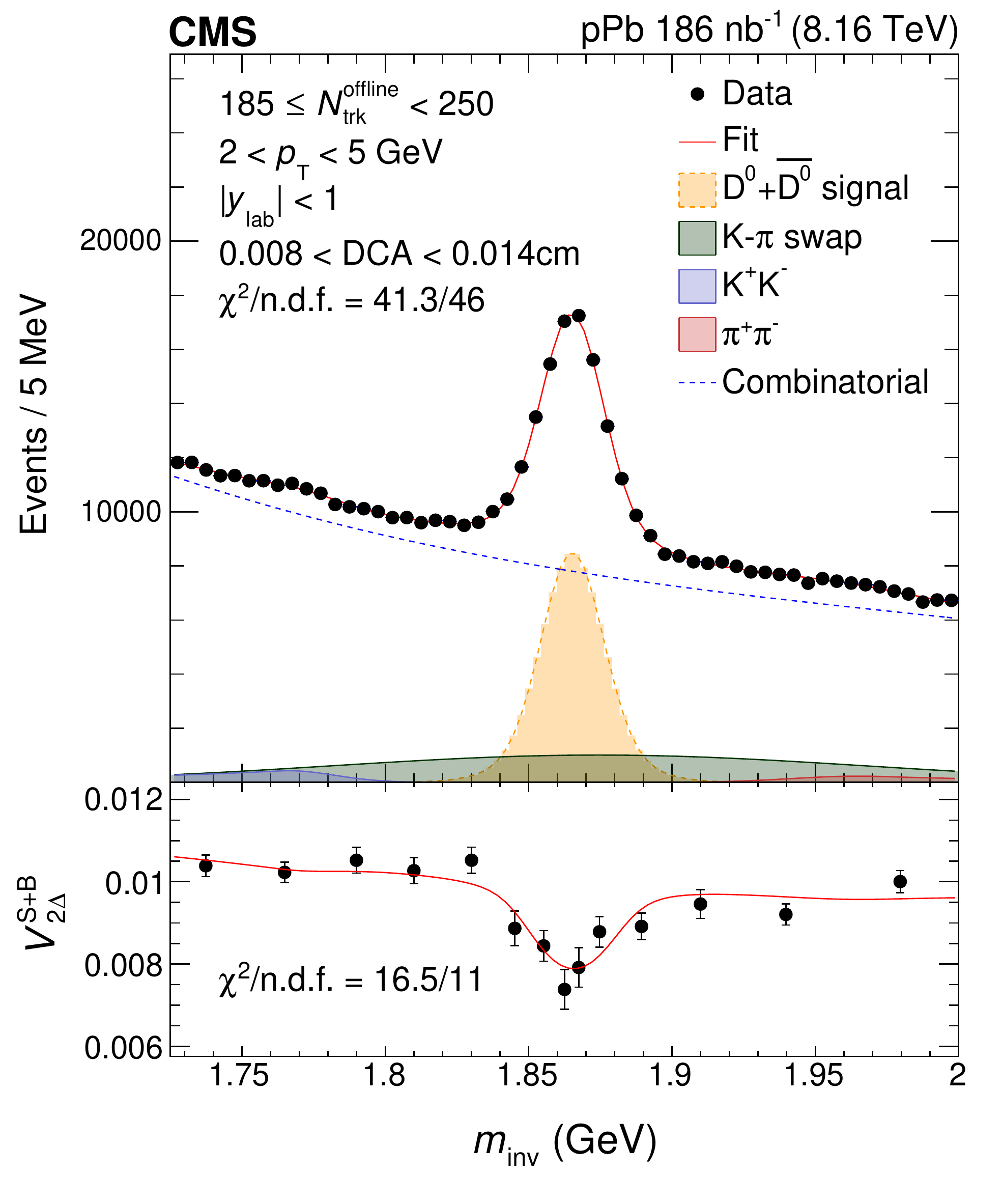}
\caption{Example of fits to the invariant mass spectrum and $V_{2\Delta}^{S+B}(m_{\text{inv}})$,
   for the BDT nonprompt-trained sample in \pPb collisions. 
   The left plot shows the fit for $\text{DCA}<0.008\cm$ and 
   the right plot is for $0.008<\text{DCA}<0.014\cm$.}
\label{fig:massfit_npd0}
\end{figure*}

\begin{figure*}[htb]
\centering
\includegraphics[width=0.35\linewidth]{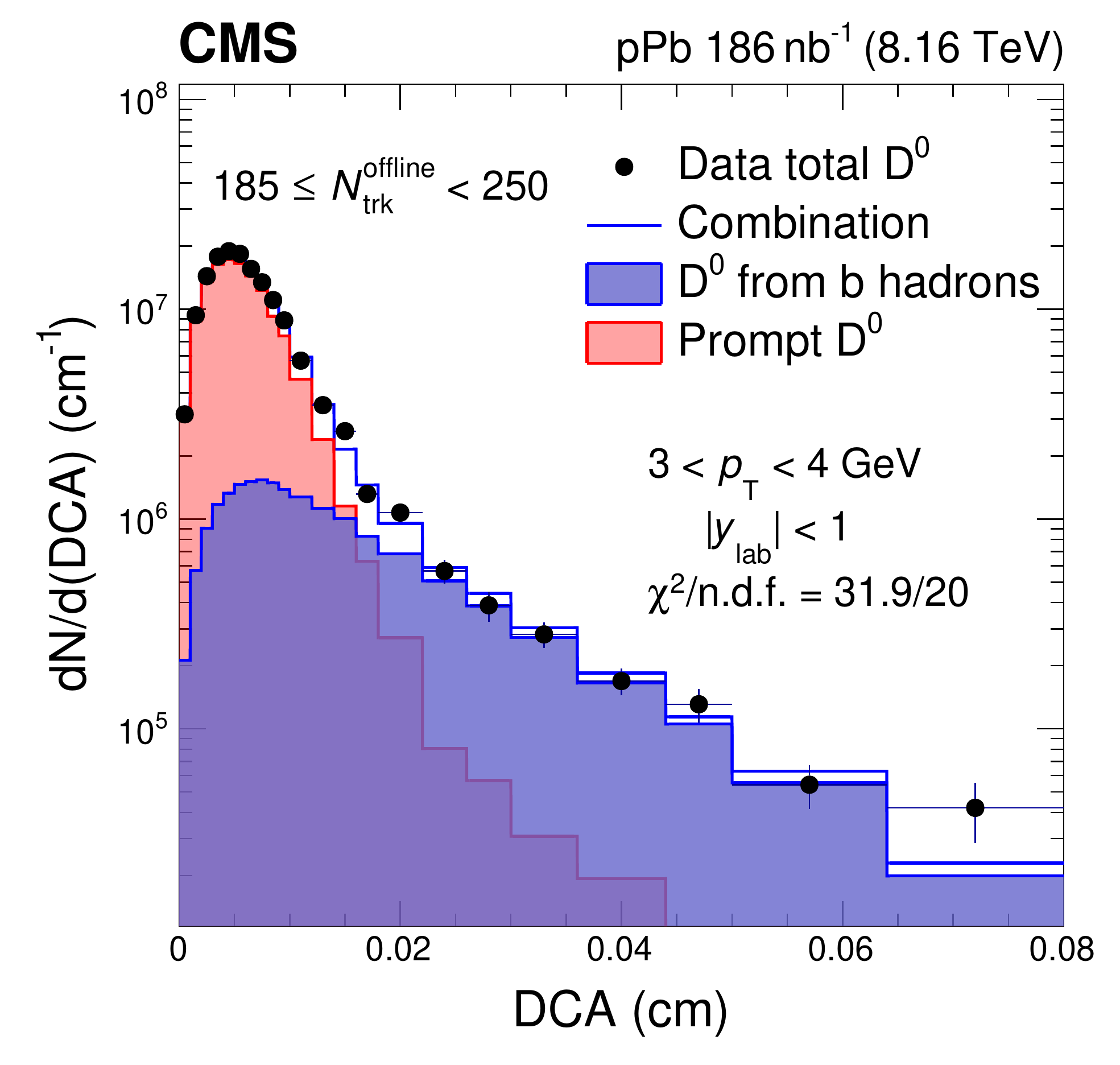}
\includegraphics[width=0.63\linewidth]{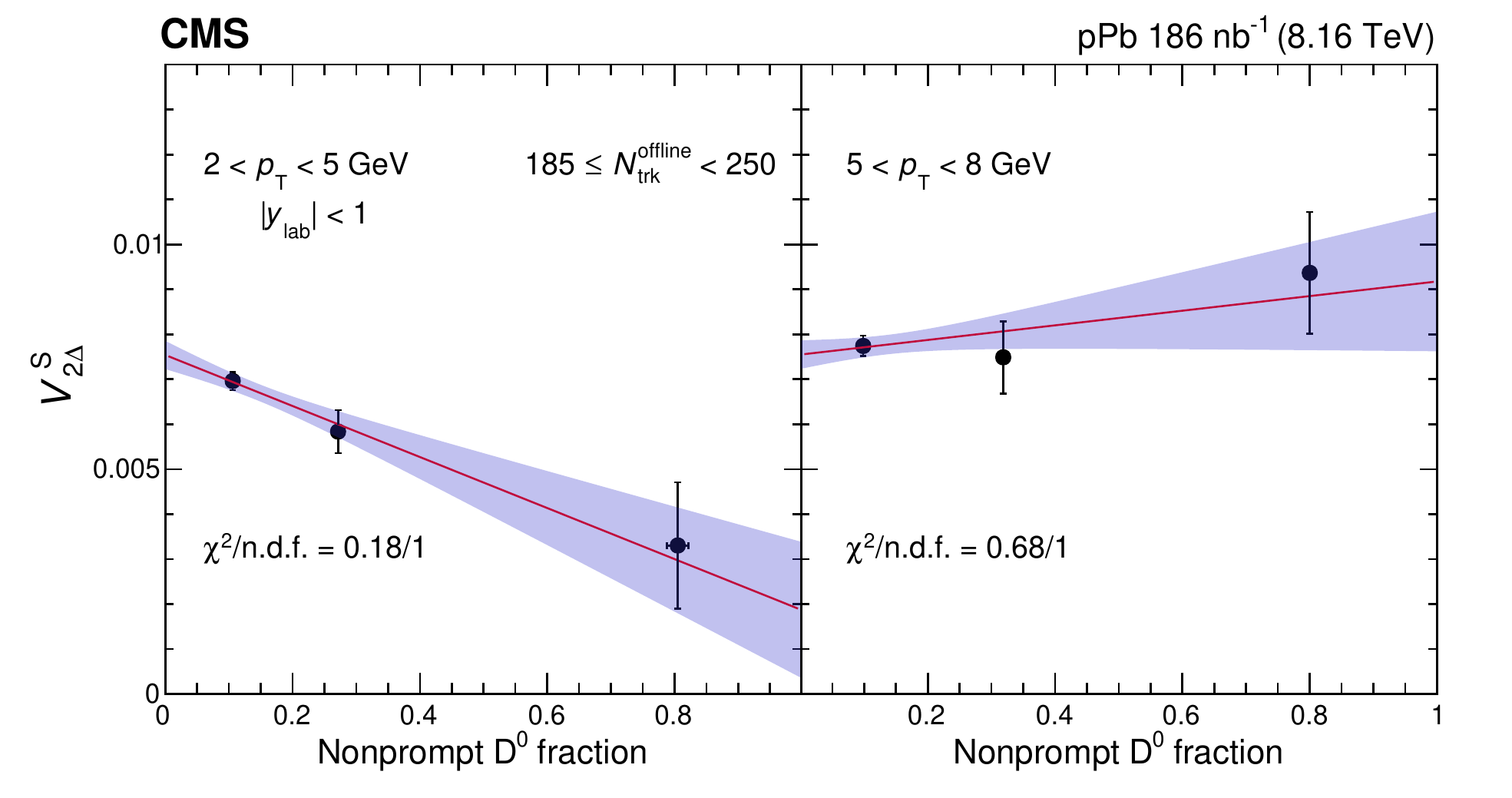}
\caption{Left: example of template fit to the \PDz meson DCA distribution in the \pt 
interval 3--4\GeV for events with $185 \leq \noff < 250$ of \pPb collisions. 
Right: inclusive \PDz $V_{2\Delta}^{S}$ values from the three DCA regions
as a function of the corresponding nonprompt \PDz fraction,
for $2<\pt<5$\GeV and $5<\pt<8$\GeV. The red line is a linear fit to $V_{2\Delta}^{S}$ data.
}
\label{fig:dca}
\end{figure*}

Figure~\ref{fig:massfitpp} shows an example of fits to the mass spectrum and $V_{2\Delta}^{S+B}(m_{\text{inv}})$,
for the BDT prompt-trained sample in the \pt interval 4--6\GeV for the multiplicity range $\noff \geq 100$
in  \pp collisions. Similar fits in \pPb data can be found in Ref.~\cite{Sirunyan:2018toe}, which are not repeated here.

For extracting the $V_{2\Delta}$ values of nonprompt \PDz mesons, the measurement and fitting procedure described
above are repeated in three separate DCA ranges, containing very different nonprompt \PDz fractions.
A linear fit by the functional form,
\begin{linenomath}
\begin{equation}
   V_{2\Delta}^{S} = f^{\text{b} \to \text{D}}V_{2\Delta}^{\text{b} \to \text{D}} 
   +(1-f^{\text{b} \to \text{D}})V_{2\Delta}^{\text{prompt D}},
\end{equation}
\end{linenomath}
\noindent to the measured \PDz $V_{2\Delta}$ values as a function of nonprompt \PDz fraction is performed
to extrapolate to the $V_{2\Delta}$ value at a nonprompt fraction of 100\%. The $f^{\text{b} \to \text{D}}$ represents
the nonprompt \PDz fraction. The $v_2$ values of nonprompt \PDz are evaluated by using Eq.~(\ref{eq:vn}).
Figure~\ref{fig:massfit_npd0} shows an example of fits to the mass spectrum and $V_{2\Delta}^{S+B}(m_{\text{inv}})$
for the BDT nonprompt-trained sample for $\text{DCA} < 0.008\cm$ and $0.008<\text{DCA}<0.014\cm$, in the \pt interval 
2--5\GeV, for the multiplicity range $185 \leq \noff < 250$ in \pPb collisions. The resulting \PDz signal $V_{2\Delta}$ 
distributions contain contributions from both prompt and nonprompt \PDz mesons.

Inclusive \PDz meson yields, extracted as a function of DCA, by fitting the invariant mass distribution in each DCA bin, 
are shown in Fig.~\ref{fig:dca} (left). A template fit to the DCA distribution is performed using template 
distributions of prompt and nonprompt \PDz mesons obtained from MC simulation to estimate
the nonprompt \PDz fractions in each of the three DCA regions used to extract inclusive \PDz $V_{2\Delta}$,
as described above. The inclusive \PDz $V_{2\Delta}$ values from the three DCA regions are then
plotted as a function of the corresponding nonprompt \PDz fraction, shown in Fig.~\ref{fig:dca} (right),
for $2<\pt<5$\GeV and $5<\pt<8$\GeV, respectively. The measurements are well described by
a linear-function fit, which is shown as a red line in Fig.~\ref{fig:dca}.

The residual contribution of back-to-back dijets to the measured $v_2$ results is corrected by subtracting correlations from
low-multiplicity events, following an identical procedure established in Refs.~\cite{Chatrchyan:2013nka,Khachatryan:2016txc}.
The Fourier coefficients, $V_{n\Delta}$, extracted from Eq.~(\ref{eq:Vn}) for $\noff < 35(20)$, in \pPb(\pp) collisions,
are subtracted from the $V_{n\Delta}$ coefficients obtained in the high-multiplicity region, with
\ifthenelse{\boolean{cms@external}}{
\begin{linenomath}
\begin{multline}
\label{eq:vnsubperiph}
V^\text{sub}_{n\Delta}=V_{n\Delta}-V_{n\Delta}(\noff<35)\\
\times\frac{N_\text{assoc}(\noff<35)}{N_\text{assoc}} \frac{Y_\text{jet}}{Y_\text{jet}(\noff<35)}.
\end{multline}
\end{linenomath}
}{
\begin{linenomath}
\begin{equation}
\label{eq:vnsubperiph}
V^\text{sub}_{n\Delta}=V_{n\Delta}-V_{n\Delta}(\noff<35)\,\frac{N_\text{assoc}(\noff<35)}{N_\text{assoc}}\,\frac{Y_\text{jet}}{Y_\text{jet}(\noff<35)}.
\end{equation}
\end{linenomath}
}
Here, $Y_\text{jet}$ represents the jet yield. It is the difference between integrals of
the short-range ($\abs{\deta}<1$) and long-range ($\abs{\deta}>2$) event-normalized 
associated yields for each multiplicity class. 
The ratio $Y_\text{jet}/Y_\text{jet}(\noff<35)$ is introduced to account
for the enhanced jet correlations resulting from the selection of higher-multiplicity
events. It is observed that the values of jet yield ratio show little
dependence on \pt over the full \pt range. For the measurement of nonprompt 
\PDz mesons, all quantities in Eq.~(\ref{eq:vnsubperiph}) are first extrapolated to 
values at a nonprompt \PDz fraction of 100\%, following the same approach as
in Fig.~\ref{fig:dca}, before applying the subtraction procedure. 
Elliptic flow ($v_2^{\text{sub}}$), corrected for residual jet correlations, is obtained from
$V^\text{sub}_{2\Delta}$
using Eq.~(\ref{eq:vn}).

\section{Systematic uncertainties}
\label{sec:systematics}

\begin{table*}[htbp]
\centering
\topcaption{Summary of systematic uncertainties on $v_2^{\text{sub}}$. 
   The ranges of systematic uncertainties correspond to the \pt ranges of \PDz mesons.
   Values are in $10^{-3}$.}
\resizebox{\linewidth}{!}{
\begin{tabular}{lccc}
Source			& Prompt \PDz in \pPb  & Nonprompt \PDz in \pPb  & Prompt \PDz in \pp \\ 
               & collisions (\ten{-3}) & collisions (\ten{-3}) & collisions (\ten{-3})\\ 
\hline
Nonprompt \PDz contamination & 3--8  &  \NA &  4--5\\
Nonprompt \PDz fraction estimation&  \NA  & 1--7 & \NA \\
Background $V_{2\Delta}$ PD	& 2--4 & 2 & 2--5\\ 
Efficiency correction		& 0.1--13 & 0.2--0.6 &0.8--13\\
Trigger bias	 & 0.6--1 & 0.1--1 & 0.4--2\\
Effect from pileup  & 2--5 & 2--5 & 4--10\\
BDT selection & 2--5 & 2 & 3--8\\
Jet subtraction & 2--7 & 14--16 & 5--49\\[\cmsTabSkip]
Total		& 5--18 & 16--17 & 13--52 \\
\end{tabular}
}
\label{tab:syst-table} 
\end{table*}

Table~\ref{tab:syst-table} summarizes the estimate of systematic uncertainties for 
the $v_2^{\text{sub}}$ of prompt and nonprompt \PDz mesons in \pPb collisions 
as well as that of prompt \PDz mesons in \pp collisions. 
The ranges of systematic uncertainties correspond to the \pt ranges of \PDz mesons.

Systematic uncertainties in the BDT selection of the \PDz candidates are evaluated by studying 
MC simulated samples. The difference between applying BDT selections and not applying those criteria is taken as
the systematic uncertainty. This procedure yields the $v_2$ uncertainties of 0.002--0.005 for prompt \PDz mesons
and 0.002 for nonprompt \PDz mesons in  \pPb collisions. In \pp collisions, it brings an uncertainty
of 0.003--0.008 on the prompt \PDz $v_2$ measurement.

Other sources of systematic uncertainty include the background mass PD, the \PDz 
meson yield correction (acceptance and efficiency correction),
the background $V_{2\Delta}$ PD, and the jet subtraction method. 
Changing the background mass PD to a second-order polynomial or an exponential function
shows negligible systematic effects.
To evaluate the uncertainties arising from the $\pt$-dependent \PDz meson yield correction,
the $v_2$ values are extracted from the corrected signal \PDz distributions and compared to the
uncorrected $v_2$ values as a conservative estimate. This
yields an uncertainty of less than 0.013. For most bins, the uncertainties from the yield correction
are less than 0.003 and are small (or negligible) compared to other sources and statistical uncertainties.
The systematic uncertainties 
from the background $v_{2}$ PD are evaluated by changing $v_{2}^{B}(m_{\text{inv}})$ 
to a second-order polynomial function of the invariant mass,
yielding an uncertainty of less than 0.005.
To study potential trigger biases, a comparison to high-multiplicity \pPb data for a given
multiplicity range that were collected using a lower threshold trigger with 100\% efficiency
is performed. The uncertainty from trigger bias is quoted as 0.001. 
Though data collected with low beam intensity are used in this analysis, there are still
additional collisions besides the one of interest per bunch crossing, which are known as pileup interactions.
The possible contamination by residual pileup interactions is also studied by 
varying the pileup selection of events in the performed analysis, from no pileup rejection at all to 
selecting events with only one reconstructed vertex. The variation of \PDz $v_2$ values
is about 0.002--0.005 in \pPb collisions, while it is about 0.004--0.010 in \pp collisions because of 
larger pileup. To study the uncertainty from jet subtraction, the ratio 
$Y_\text{jet}$/$Y_\text{jet}(\noff<35)$ is varied by
one standard deviation. It yields an uncertainty of 0.002--0.007 for prompt \PDz mesons and
0.016--0.017 for nonprompt \PDz in \pPb collisions. In \pp collisions, it yields an uncertainty of
0.013--0.049 for prompt \PDz mesons. This effect diminishes towards high multiplicity regions
because of the small \noff ratio according to Eq.~(\ref{eq:vnsubperiph}).

For the measurement of prompt \PDz mesons, the contribution from nonprompt \PDz 
mesons is significantly suppressed. No explicit correction is applied and a systematic uncertainty 
is quoted instead. Based on the prediction for AA collisions that \PB\ mesons have a smaller $v_2$
than light-flavor particles because of the larger mass of the \PQb quark~\cite{Nahrgang:2014vza,He:2014cla,Xu:2015bbz},
the nonprompt \PDz $v_2$ values are assumed to lie between 0 and those of strange hadrons.
The $v_2$ for prompt \PDz is thus reestimated with the bounds of nonprompt \PDz $v_2$ and the extracted nonprompt \PDz fractions
and the change in $v_{2}$ signal is found to be smaller than 0.008.
For the measurement of nonprompt \PDz mesons, a major systematic uncertainty comes
from the determination of nonprompt \PDz fraction in different DCA regions. 
The DCA template distributions of prompt and nonprompt \PDz mesons from MC simulation are smeared via scaling the width of
these distributions. The variation of DCA width is 2--8\%, based on
the best $\chi^2$ fit to data. The resulting variation in
the extracted nonprompt \PDz $v_2$ are quoted as a systematic uncertainty of 0.007.

All sources of systematic uncertainties are added in quadrature to obtain the total systematic uncertainty.
The total systematic uncertainties for prompt and nonprompt \PDz mesons in \pPb collisions 
yield 0.005--0.018 and 0.016--0.017, respectively.
For prompt \PDz mesons in \pp collisions, the total systematic uncertainties are quoted as 0.013--0.052.

\section{Results}
\label{sec:results}

\begin{figure}[htb]
\centering
\includegraphics[width=\cmsFigWidth]{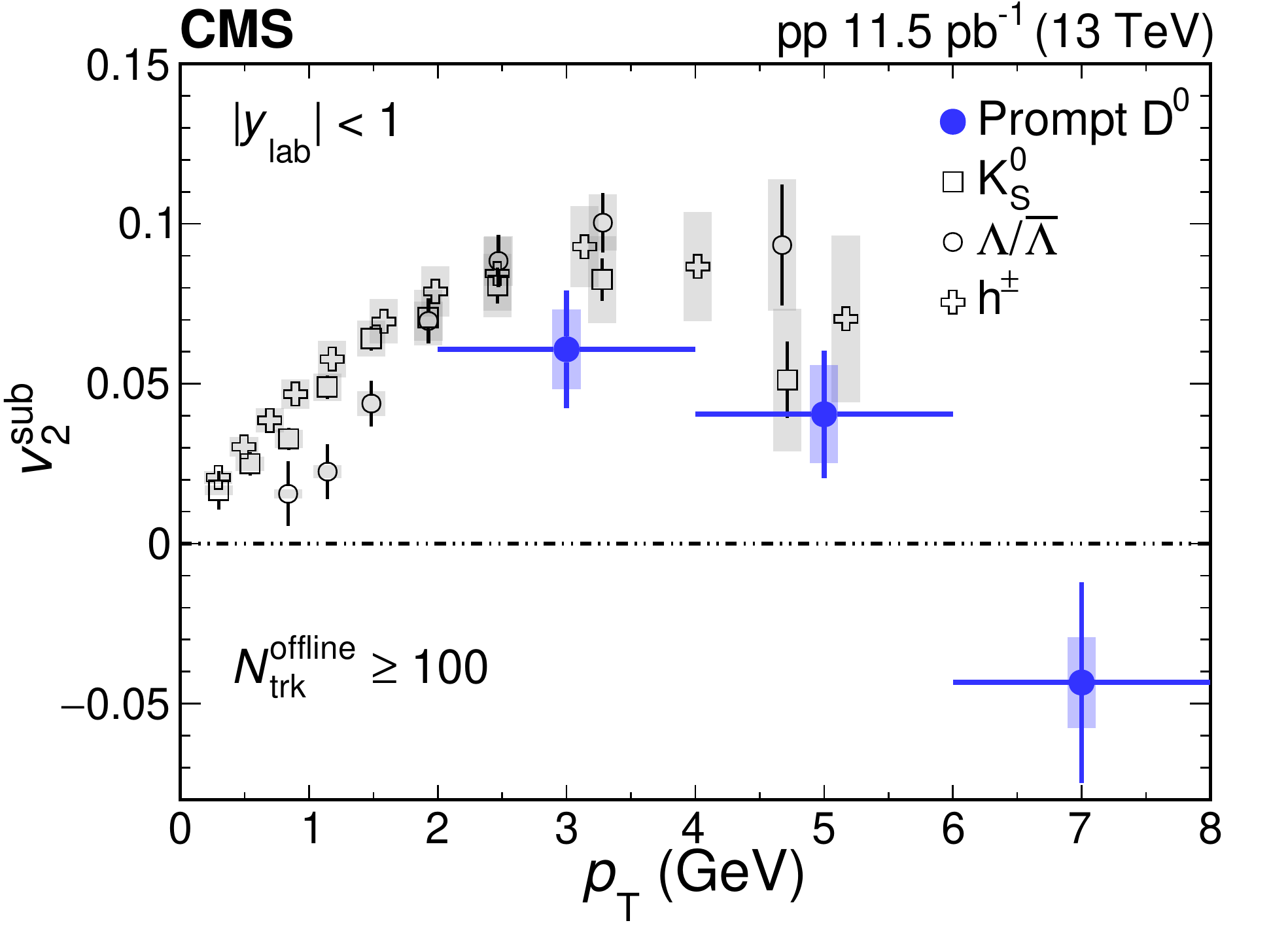}
   \caption{Results of $v_2^{\text{sub}}$ for prompt \PDz mesons, as
   a function of \pt for $\abs{y_{\text{lab}}}<1$, with $\noff \geq 100$ in  \pp collisions at $\roots  = 13\TeV$.
Published data for charged particles, \PKzS mesons and \PgL baryons are also shown for comparison~\cite{Khachatryan:2016txc}.
The vertical bars correspond to the statistical uncertainties, while the shaded areas denote
the systematic uncertainties. The horizontal bars represent the width of the \pt bins.}
\label{fig:v2pp}
\end{figure}

The $v_{2}^{\text{sub}}$ results of prompt \PDz mesons in  \pp collisions at $\roots\ = 13\TeV$, 
are presented in Fig.~\ref{fig:v2pp} as a function of \pt for $\abs{y_{\text{lab}}}<1$, with $\noff \geq 100$.
Published data 
for light-flavor hadrons including inclusive charged particles (dominated by pions), \PKzS mesons
and \PgL baryons are also shown for comparison~\cite{Khachatryan:2016txc}. 
The positive
$v_2$ signal ($0.061\pm0.018 \stat \pm0.013 \syst$) 
over a \pt range of $\sim$2--4\GeV for prompt charm hadrons provides indications 
of the collectivity of charm quarks in \pp collisions,
with a declining trend toward higher \pt. The $v_2$ magnitude for prompt \PDz mesons
is found to be compatible with light-flavor hadron species, though slightly smaller by about one standard deviation.
The results suggest that collectivity is being developed for charm hadrons in  \pp collisions,
comparable (or slightly weaker) than that for light-flavor hadrons.
This finding is similar to the observation made in \pPb collisions at $\sqrtsNN = 8.16\TeV$ over a similar \pt range 
at higher multiplicities $185 \leq \noff < 250$~\cite{Sirunyan:2018toe}.

\begin{figure}[htb!]
\centering
\includegraphics[width=\cmsFigWidth]{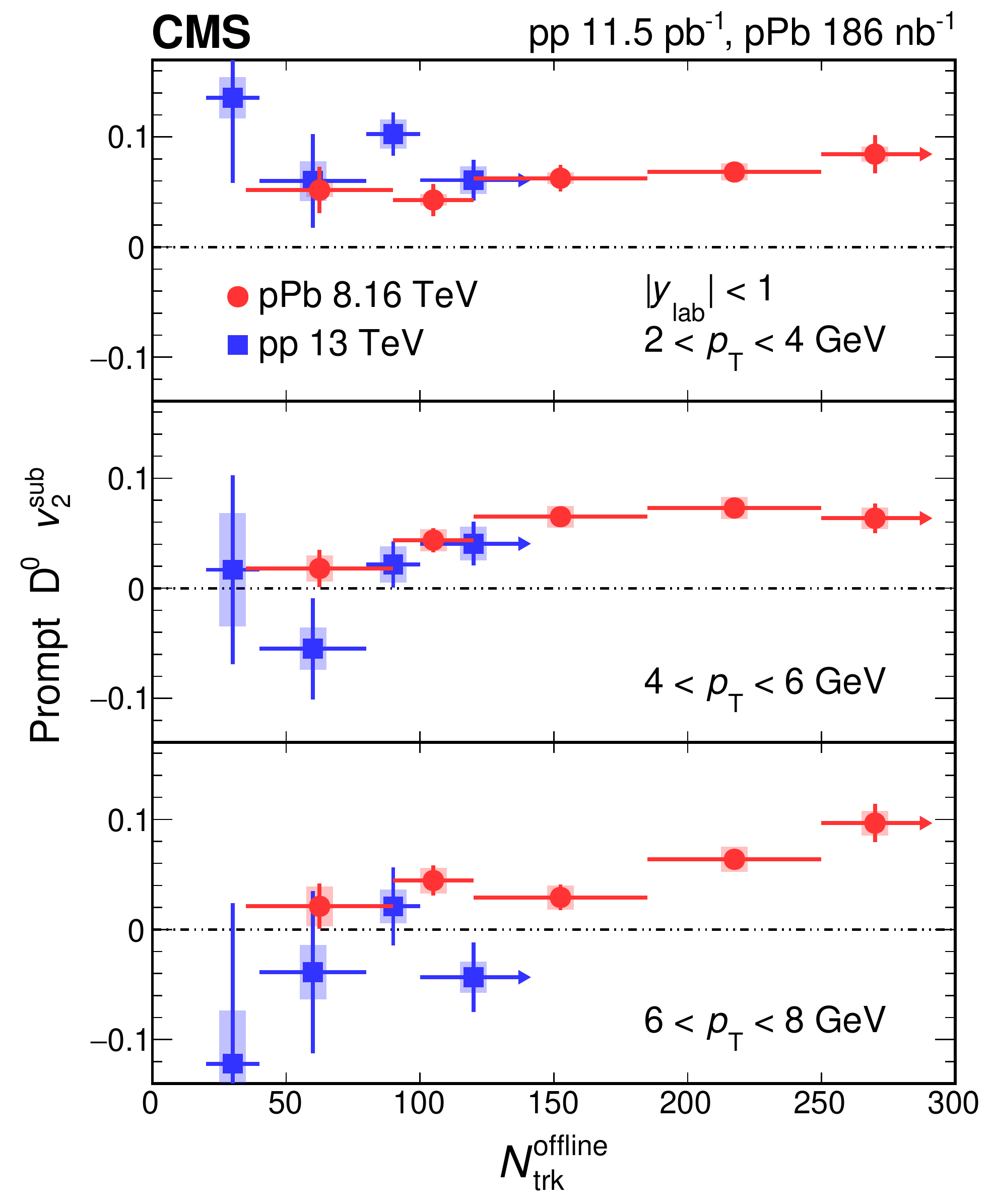}
   \caption{Results of $v_2^{\text{sub}}$ for prompt \PDz mesons, as
   a function of event multiplicity for three different \pt ranges, with $\abs{y_{\text{lab}}}<1$ in 
 \pp collisions at $\roots\ = 13$\TeV\ and \pPb collisions at $\sqrtsNN  = 8.16$\TeV.
The vertical bars correspond to statistical uncertainties, while the shaded areas denote
the systematic uncertainties. 
The y-axis is zoomed in to better display the data; the uncertainties are symmetric with respect to their central values.
The horizontal bars represent the width of the \noff bins.
The right-most points with right-hand arrows
correspond to $\noff\geq 100$ for \pp collisions and 
   $\noff\geq250$ for \pPb collisions. The $v_2^{\text{sub}}$ values in \pPb collisions 
with $185\leq\noff < 250$ are measured in different \pt ranges from Ref.~\cite{Sirunyan:2018toe}
   and are found to be consistent with Ref.~\cite{Sirunyan:2018toe}.
}
\label{fig:v2ntrk}
\end{figure}

To further investigate possible system size dependence of collectivity for charm hadrons in small colliding systems,
$v_2$ for prompt \PDz mesons in \pPb and \pp collisions are both measured in different multiplicity classes. 
The prompt \PDz $v_2$
as a function of event multiplicity for three different \pt ranges: $2<\pt<4\GeV$, $4<\pt<6\GeV$, 
and $6<\pt<8\GeV$ are presented in Fig.~\ref{fig:v2ntrk}. 
At similar multiplicities of $\noff \sim 100$, the prompt \PDz $v_2$ values are found to be comparable
within uncertainties in  \pp and \pPb systems. For $2<\pt<4\GeV$, the measured results of prompt \PDz
provide indications of positive $v_2$ down to $\noff \sim 50$ with a significance of
more than $2.4$ standard deviations in \pPb collisions, while for $6<\pt<8\GeV$
the prompt \PDz $v_2$ signal tends to diminish in the low multiplicity regions.
No clear multiplicity dependence can be determined for \pp data, because of large statistical 
uncertainties at low multiplicities.

\begin{figure}[htb!]
\centering
\includegraphics[width=\cmsFigWidth]{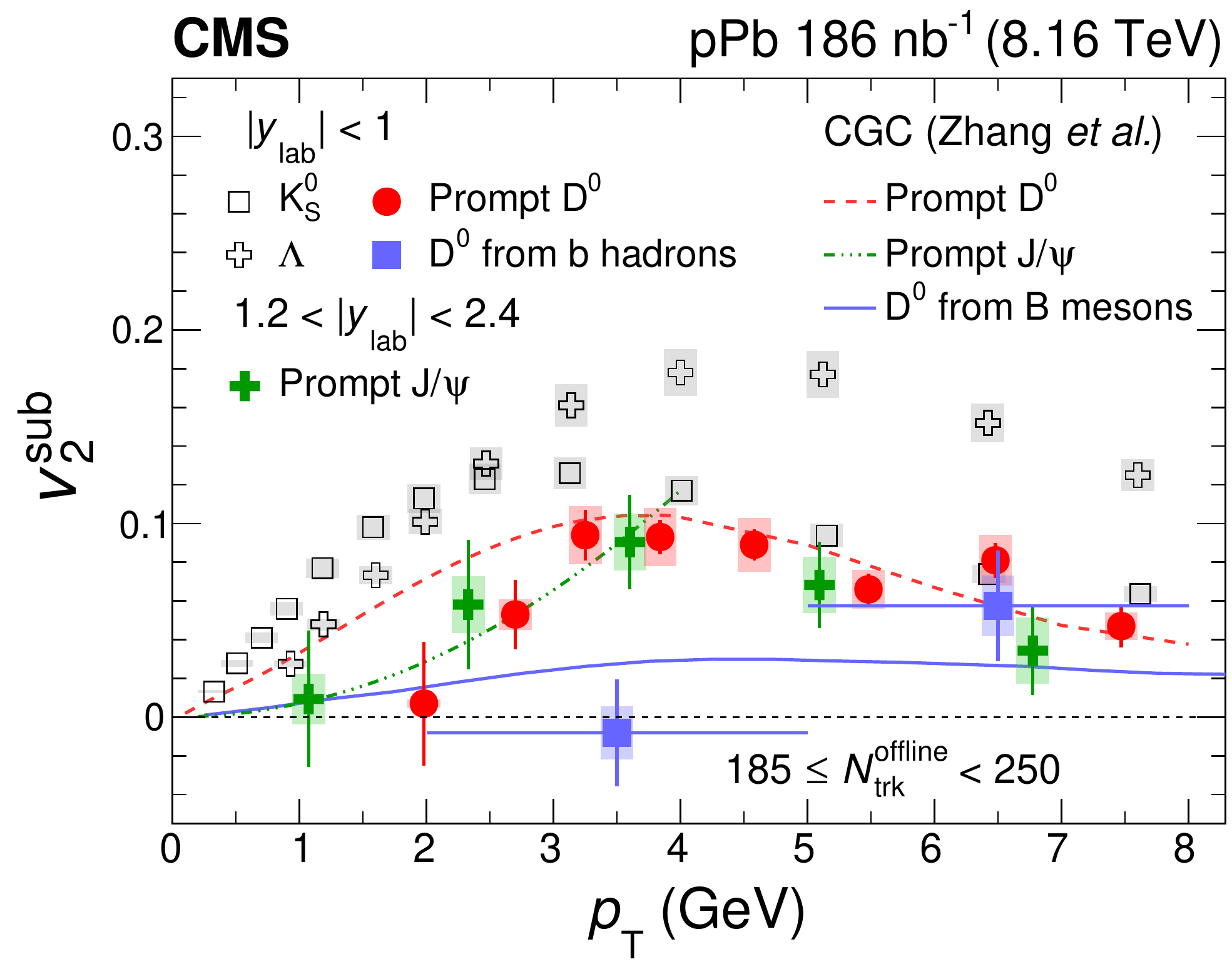}
   \caption{Results of $v_2^{\text{sub}}$ for prompt and nonprompt \PDz mesons, as well as \PKzS mesons,
   \PgL baryons for $\abs{y_{\text{lab}}}<1$, and prompt \PJGy\ mesons for $1.2<\abs{y_{\text{lab}}}<2.4$, as functions of \pt
with $185 \leq \noff < 250$ in \pPb  collisions at $\sqrtsNN  = 8.16\TeV$ \cite{Sirunyan:2018toe,Sirunyan:2018kiz}.
The vertical bars correspond to statistical uncertainties, while the shaded areas denote
the systematic uncertainties. The horizontal bars represent the width of the nonprompt \PDz \pt bins.
The dashed, dash-dotted, and solid lines, show the theoretical calculations of prompt \PDz, \PJGy, and nonprompt \PDz
mesons, respectively, within the CGC framework~\cite{Zhang:2019dth,Zhang:2020ayy}.
 }
\label{fig:v2np}
\end{figure}

The $v_2^{\text{sub}}$ results for nonprompt \PDz mesons from beauty hadron decays are shown in Fig.~\ref{fig:v2np} as a function of
\pt for \pPb collisions at 8.16\TeV\ with $185 \leq \noff < 250$. 
The extracted $v_{2}^{\text{sub}}$ values are $-0.008\pm 0.028 \stat \pm 0.016 \syst$  
for $2<\pt<5\GeV$ and $0.057\pm 0.029 \stat \pm 0.017 \syst$ for $5<\pt<8\GeV$.
At low \pt, the nonprompt \PDz $v_2$ is consistent with zero, while at high \pt, 
a hint of a positive $v_2$ value for beauty mesons is suggested but not significant within statistical and systematic uncertainties.
Previously published $v_2$ data for prompt \PDz mesons and strange hadrons are also shown~\cite{Sirunyan:2018toe}.

At $\pt \sim 2$--5\GeV, the nonprompt \PDz meson $v_2$ from beauty hadron decays is observed to be
smaller than that for prompt \PDz mesons with a significance of 2.7 standard deviations.
Based on MC simulations with \EVTGEN and \PYTHIA~\cite{Lange:2001uf,Sjostrand:2014zea},
nonprompt \PDz mesons carry more than 50\% of \PB transverse momenta.
The deviation of nonprompt \PDz meson azimuthal distributions from \PB mesons
could reduce the extracted $v_2$ values at fixed \PB meson \pt.
Taking the gluon saturation model as an example, the maximum $v_2$ value of \PB mesons is 
at $\pt \sim 6 \GeV$~\cite{Zhang:2020ayy}, while 
the maximum $v_2$ value of nonprompt \PDz mesons is about 70\% of that of \PB mesons
at \PDz $\pt \sim 4 \GeV$ due to the effects discussed above.
These studies suggest a flavor hierarchy of the collectivity signal that tends to
diminish for the heavier beauty hadrons.
This is qualitatively consistent with the scenario of $v_2$ being generated via final-state
rescatterings, where heavier quarks tend to develop a weaker collective $v_2$ signal~\cite{Dong:2019byy}.
The ordering of muon $v_2$ from charm and beauty decay at low \pt is also observed
in \PbPb collisions where final-state scatterings play an important role~\cite{Aad:2020grf}.

Correlations at the initial stage of the collision 
between partons originating from projectile protons and dense gluons in the lead nucleus 
are able to generate sizable elliptic flow in the color glass condensate (CGC) framework~\cite{Dusling:2015gta,Zhang:2019dth,Zhang:2020ayy}.
These CGC calculations of $v_2$ signals for prompt \PJGy mesons, as well as 
prompt and nonprompt (from \PB meson decay) \PDz mesons, are compared with data in Fig.~\ref{fig:v2np}.
The qualitative agreement between data and theory suggests that 
initial-state effects may play an important role in the generation of collectivity 
for these particles in \pPb collisions. The CGC framework also predicts a flavor hierarchy 
between prompt and nonprompt \PDz for $\pt \sim 2$--5\GeV, again consistent with the data within uncertainties.

\section{Summary}
\label{sec:summary}

The first measurements of elliptic azimuthal anisotropies for prompt \PDz mesons
in proton-proton (\pp) collisions at center-of-mass energy $\roots = 13\TeV$, and for nonprompt \PDz mesons from 
beauty hadron decays in proton-lead (\pPb) collisions at nucleon-nucleon center-of-mass energy $\sqrtsNN = 8.16\TeV$ are presented. 
In  \pp collisions with multiplicities of $\noff\geq100$, the second Fourier harmonic coefficient ($v_2$) of the azimuthal distributions
for prompt \PDz mesons are measured over the transverse momentum (\pt) range of 2--8\GeV,
with indications of positive $v_2$ signals over the
\pt range of 2--4\GeV. These values are found to be comparable 
(or slightly smaller) to those of light-flavor 
hadron species. At similar event multiplicities, the prompt \PDz meson
$v_2$ signals in  \pp and \pPb collisions are found to be comparable in magnitude. The $v_2$ values
of open beauty hadrons are extracted for the first time via non-prompt \PDz mesons in \pPb
collisions, with magnitudes smaller than those for prompt \PDz mesons for $\pt \sim 2$--5\GeV.
The new measurements of charm hadron $v_2$ in the \pp system and the indications of mass dependence of
heavy-flavor hadron $v_2$ in the \pPb system provide insights into the origin of heavy-flavor
quark collectivity in small colliding systems.

\begin{acknowledgments}
   We congratulate our colleagues in the CERN accelerator departments for the excellent performance of the LHC and thank the technical and administrative staffs at CERN and at other CMS institutes for their contributions to the success of the CMS effort. In addition, we gratefully acknowledge the computing centers and personnel of the Worldwide LHC Computing Grid for delivering so effectively the computing infrastructure essential to our analyses. Finally, we acknowledge the enduring support for the construction and operation of the LHC and the CMS detector provided by the following funding agencies: BMBWF and FWF (Austria); FNRS and FWO (Belgium); CNPq, CAPES, FAPERJ, FAPERGS, and FAPESP (Brazil); MES (Bulgaria); CERN; CAS, MoST, and NSFC (China); COLCIENCIAS (Colombia); MSES and CSF (Croatia); RIF (Cyprus); SENESCYT (Ecuador); MoER, ERC IUT, PUT and ERDF (Estonia); Academy of Finland, MEC, and HIP (Finland); CEA and CNRS/IN2P3 (France); BMBF, DFG, and HGF (Germany); GSRT (Greece); NKFIA (Hungary); DAE and DST (India); IPM (Iran); SFI (Ireland); INFN (Italy); MSIP and NRF (Republic of Korea); MES (Latvia); LAS (Lithuania); MOE and UM (Malaysia); BUAP, CINVESTAV, CONACYT, LNS, SEP, and UASLP-FAI (Mexico); MOS (Montenegro); MBIE (New Zealand); PAEC (Pakistan); MSHE and NSC (Poland); FCT (Portugal); JINR (Dubna); MON, RosAtom, RAS, RFBR, and NRC KI (Russia); MESTD (Serbia); SEIDI, CPAN, PCTI, and FEDER (Spain); MOSTR (Sri Lanka); Swiss Funding Agencies (Switzerland); MST (Taipei); ThEPCenter, IPST, STAR, and NSTDA (Thailand); TUBITAK and TAEK (Turkey); NASU (Ukraine); STFC (United Kingdom); DOE and NSF (USA).
   
   \hyphenation{Rachada-pisek} Individuals have received support from the Marie-Curie program and the European Research Council and Horizon 2020 Grant, contract Nos.\ 675440, 752730, and 765710 (European Union); the Leventis Foundation; the A.P.\ Sloan Foundation; the Alexander von Humboldt Foundation; the Belgian Federal Science Policy Office; the Fonds pour la Formation \`a la Recherche dans l'Industrie et dans l'Agriculture (FRIA-Belgium); the Agentschap voor Innovatie door Wetenschap en Technologie (IWT-Belgium); the F.R.S.-FNRS and FWO (Belgium) under the ``Excellence of Science -- EOS" -- be.h project n.\ 30820817; the Beijing Municipal Science \& Technology Commission, No. Z191100007219010; the Ministry of Education, Youth and Sports (MEYS) of the Czech Republic; the Deutsche Forschungsgemeinschaft (DFG) under Germany's Excellence Strategy -- EXC 2121 ``Quantum Universe" -- 390833306; the Lend\"ulet (``Momentum") Program and the J\'anos Bolyai Research Scholarship of the Hungarian Academy of Sciences, the New National Excellence Program \'UNKP, the NKFIA research grants 123842, 123959, 124845, 124850, 125105, 128713, 128786, and 129058 (Hungary); the Council of Science and Industrial Research, India; the HOMING PLUS program of the Foundation for Polish Science, cofinanced from European Union, Regional Development Fund, the Mobility Plus program of the Ministry of Science and Higher Education, the National Science Center (Poland), contracts Harmonia 2014/14/M/ST2/00428, Opus 2014/13/B/ST2/02543, 2014/15/B/ST2/03998, and 2015/19/B/ST2/02861, Sonata-bis 2012/07/E/ST2/01406; the National Priorities Research Program by Qatar National Research Fund; the Ministry of Science and Higher Education, project no. 02.a03.21.0005 (Russia); the Tomsk Polytechnic University Competitiveness Enhancement Program; the Programa Estatal de Fomento de la Investigaci{\'o}n Cient{\'i}fica y T{\'e}cnica de Excelencia Mar\'{\i}a de Maeztu, grant MDM-2015-0509 and the Programa Severo Ochoa del Principado de Asturias; the Thalis and Aristeia programs cofinanced by EU-ESF and the Greek NSRF; the Rachadapisek Sompot Fund for Postdoctoral Fellowship, Chulalongkorn University and the Chulalongkorn Academic into Its 2nd Century Project Advancement Project (Thailand); the Kavli Foundation; the Nvidia Corporation; the SuperMicro Corporation; the Welch Foundation, contract C-1845; and the Weston Havens Foundation (USA).\end{acknowledgments}

\bibliography{auto_generated}

\providecommand{\href}[2]{#2}\begingroup\raggedright\begin{thebibliography}{10}%
\makeatletter
\providecommand{\hrefCMSnoop }[0]{\@secondoftwo}%
\makeatother
\providecommand{\doi}{\texttt{doi:}\begingroup \urlstyle{tt}\Url}

\bibitem{Ackermann:2000tr}
\hrefCMSnoop {}{{STAR} Collaboration, ``{Elliptic flow in Au+Au collisions at
  $\sqrtsNN = 130 \GeV$}'',} \textit{ Phys. Rev. Lett.} \textbf{ 86} (2001)
  402,
  \href{http://dx.doi.org/10.1103/PhysRevLett.86.402}{\doi{10.1103/PhysRevLett.86.402}},
\href{http://www.arXiv.org/abs/nucl-ex/0009011}{\texttt{arXiv:nucl-ex/0009011}}.

\bibitem{Adler:2003kt}
\hrefCMSnoop {}{{PHENIX} Collaboration, ``{Elliptic flow of identified hadrons
  in Au+Au collisions at $\sqrtsNN = 200\GeV$}'',} \textit{ Phys. Rev. Lett.}
  \textbf{ 91} (2003) 182301,
  \href{http://dx.doi.org/10.1103/PhysRevLett.91.182301}{\doi{10.1103/PhysRevLett.91.182301}},
\href{http://www.arXiv.org/abs/nucl-ex/0305013}{\texttt{arXiv:nucl-ex/0305013}}.

\bibitem{Adams:2005ph}
\hrefCMSnoop {}{{STAR} Collaboration, ``{Distributions of charged hadrons
  associated with high transverse momentum particles in pp and Au+Au collisions
  at $\sqrtsNN = 200$\GeV}'',} \textit{ Phys. Rev. Lett.} \textbf{ 95} (2005)
  152301,
  \href{http://dx.doi.org/10.1103/PhysRevLett.95.152301}{\doi{10.1103/PhysRevLett.95.152301}},
\href{http://www.arXiv.org/abs/nucl-ex/0501016}{\texttt{arXiv:nucl-ex/0501016}}.

\bibitem{Alver:2008gk}
\hrefCMSnoop {}{{PHOBOS} Collaboration, ``{System size dependence of cluster
  properties from two- particle angular correlations in Cu+Cu and Au+Au
  collisions at $\sqrtsNN = 200$\GeV}'',} \textit{ Phys. Rev. C} \textbf{ 81}
  (2010) 024904,
  \href{http://dx.doi.org/10.1103/PhysRevC.81.024904}{\doi{10.1103/PhysRevC.81.024904}},
\href{http://www.arXiv.org/abs/0812.1172}{\texttt{arXiv:0812.1172}}.

\bibitem{Aamodt:2010pa}
\hrefCMSnoop {}{{ALICE Collaboration}, ``{Elliptic flow of charged particles in
  Pb-Pb collisions at 2.76 TeV}'',} \textit{ Phys. Rev. Lett.} \textbf{ 105}
  (2010) 252302,
  \href{http://dx.doi.org/10.1103/PhysRevLett.105.252302}{\doi{10.1103/PhysRevLett.105.252302}},
\href{http://www.arXiv.org/abs/1011.3914}{\texttt{arXiv:1011.3914}}.

\bibitem{Chatrchyan:2012ta}
\hrefCMSnoop {}{{CMS Collaboration}, ``{Measurement of the elliptic anisotropy
  of charged particles produced in PbPb collisions at nucleon-nucleon
  center-of-mass energy = 2.76 TeV}'',} \textit{ Phys. Rev. C} \textbf{ 87}
  (2013) 014902,
  \href{http://dx.doi.org/10.1103/PhysRevC.87.014902}{\doi{10.1103/PhysRevC.87.014902}},
\href{http://www.arXiv.org/abs/1204.1409}{\texttt{arXiv:1204.1409}}.

\bibitem{Ollitrault:1992bk}
\hrefCMSnoop {}{J.-Y. Ollitrault, ``{Anisotropy as a signature of transverse
  collective flow}'',} \textit{ Phys. Rev. D} \textbf{ 46} (1992) 229,
\href{http://dx.doi.org/10.1103/PhysRevD.46.229}{\doi{10.1103/PhysRevD.46.229}}.

\bibitem{Heinz:2013th}
\hrefCMSnoop {}{U.~Heinz and R.~Snellings, ``{Collective flow and viscosity in
  relativistic heavy-ion collisions}'',} \textit{ Ann. Rev. Nucl. Part. Sci.}
  \textbf{ 63} (2013) 123,
  \href{http://dx.doi.org/10.1146/annurev-nucl-102212-170540}{\doi{10.1146/annurev-nucl-102212-170540}},
\href{http://www.arXiv.org/abs/1301.2826}{\texttt{arXiv:1301.2826}}.

\bibitem{Gale:2013da}
\hrefCMSnoop {}{C.~Gale, S.~Jeon, and B.~Schenke, ``{Hydrodynamic modeling of
  heavy-ion collisions}'',} \textit{ Int. J. Mod. Phys. A} \textbf{ 28} (2013)
  1340011,
  \href{http://dx.doi.org/10.1142/S0217751X13400113}{\doi{10.1142/S0217751X13400113}},
\href{http://www.arXiv.org/abs/1301.5893}{\texttt{arXiv:1301.5893}}.

\bibitem{Alver:2009id}
\hrefCMSnoop {}{{PHOBOS} Collaboration, ``{High transverse momentum triggered
  correlations over a large pseudorapidity acceptance in Au+Au collisions at
  $\sqrtsNN = 200$\GeV}'',} \textit{ Phys. Rev. Lett.} \textbf{ 104} (2010)
  062301,
  \href{http://dx.doi.org/10.1103/PhysRevLett.104.062301}{\doi{10.1103/PhysRevLett.104.062301}},
\href{http://www.arXiv.org/abs/0903.2811}{\texttt{arXiv:0903.2811}}.

\bibitem{Abelev:2009af}
\hrefCMSnoop {}{{STAR} Collaboration, ``{Long range rapidity correlations and
  jet production in high energy nuclear collisions}'',} \textit{ Phys. Rev. C}
  \textbf{ 80} (2009) 064912,
  \href{http://dx.doi.org/10.1103/PhysRevC.80.064912}{\doi{10.1103/PhysRevC.80.064912}},
\href{http://www.arXiv.org/abs/0909.0191}{\texttt{arXiv:0909.0191}}.

\bibitem{Chatrchyan:2011eka}
\hrefCMSnoop {}{{CMS Collaboration}, ``{Long-range and short-range dihadron
  angular correlations in central PbPb collisions at a nucleon-nucleon center
  of mass energy of 2.76 TeV}'',} \textit{ JHEP} \textbf{ 07} (2011) 076,
  \href{http://dx.doi.org/10.1007/JHEP07(2011)076}{\doi{10.1007/JHEP07(2011)076}},
\href{http://www.arXiv.org/abs/1105.2438}{\texttt{arXiv:1105.2438}}.

\bibitem{Chatrchyan:2012wg}
\hrefCMSnoop {}{{CMS Collaboration}, ``{Centrality dependence of dihadron
  correlations and azimuthal anisotropy harmonics in PbPb collisions at
  $\sqrtsNN = 2.76$\TeV}'',} \textit{ Eur. Phys. J. C} \textbf{ 72} (2012)
  2012,
  \href{http://dx.doi.org/10.1140/epjc/s10052-012-2012-3}{\doi{10.1140/epjc/s10052-012-2012-3}},
\href{http://www.arXiv.org/abs/1201.3158}{\texttt{arXiv:1201.3158}}.

\bibitem{ATLAS:2012at}
\hrefCMSnoop {}{{ATLAS Collaboration}, ``{Measurement of the azimuthal
  anisotropy for charged particle production in $\sqrtsNN = 2.76$\TeV lead-lead
  collisions with the ATLAS detector}'',} \textit{ Phys. Rev. C} \textbf{ 86}
  (2012) 014907,
  \href{http://dx.doi.org/10.1103/PhysRevC.86.014907}{\doi{10.1103/PhysRevC.86.014907}},
\href{http://www.arXiv.org/abs/1203.3087}{\texttt{arXiv:1203.3087}}.

\bibitem{CMS:2013bza}
\hrefCMSnoop {}{{CMS Collaboration}, ``{Studies of azimuthal dihadron
  correlations in ultra-central PbPb collisions at $\sqrtsNN = 2.76\TeV$}'',}
  \textit{ JHEP} \textbf{ 02} (2014) 088,
  \href{http://dx.doi.org/10.1007/JHEP02(2014)088}{\doi{10.1007/JHEP02(2014)088}},
\href{http://www.arXiv.org/abs/1312.1845}{\texttt{arXiv:1312.1845}}.

\bibitem{Khachatryan:2010gv}
\hrefCMSnoop {}{{CMS Collaboration}, ``{Observation of long-range near-side
  angular correlations in proton-proton collisions at the LHC}'',} \textit{
  JHEP} \textbf{ 09} (2010) 091,
  \href{http://dx.doi.org/10.1007/JHEP09(2010)091}{\doi{10.1007/JHEP09(2010)091}},
\href{http://www.arXiv.org/abs/1009.4122}{\texttt{arXiv:1009.4122}}.

\bibitem{Aad:2015gqa}
\hrefCMSnoop {}{{ATLAS Collaboration}, ``{Observation of long-range elliptic
  azimuthal anisotropies in $\roots=$13 and 2.76 TeV pp collisions with the
  ATLAS detector}'',} \textit{ Phys. Rev. Lett.} \textbf{ 116} (2016) 172301,
  \href{http://dx.doi.org/10.1103/PhysRevLett.116.172301}{\doi{10.1103/PhysRevLett.116.172301}},
\href{http://www.arXiv.org/abs/1509.04776}{\texttt{arXiv:1509.04776}}.

\bibitem{Khachatryan:2015lva}
\hrefCMSnoop {}{{CMS Collaboration}, ``{Measurement of long-range near-side
  two-particle angular correlations in pp collisions at $\roots =$13 TeV}'',}
  \textit{ Phys. Rev. Lett.} \textbf{ 116} (2016) 172302,
  \href{http://dx.doi.org/10.1103/PhysRevLett.116.172302}{\doi{10.1103/PhysRevLett.116.172302}},
\href{http://www.arXiv.org/abs/1510.03068}{\texttt{arXiv:1510.03068}}.

\bibitem{Khachatryan:2016txc}
\hrefCMSnoop {}{{CMS Collaboration}, ``{Evidence for collectivity in pp
  collisions at the LHC}'',} \textit{ Phys. Lett. B} \textbf{ 765} (2017) 193,
  \href{http://dx.doi.org/10.1016/j.physletb.2016.12.009}{\doi{10.1016/j.physletb.2016.12.009}},
\href{http://www.arXiv.org/abs/1606.06198}{\texttt{arXiv:1606.06198}}.

\bibitem{Aad:2019aol}
\hrefCMSnoop {}{{ATLAS Collaboration}, ``{Measurement of azimuthal anisotropy
  of muons from charm and bottom hadrons in pp collisions at $\sqrt{s}=13$ TeV
  with the ATLAS detector}'',} \textit{ Phys. Rev. Lett.} \textbf{ 124} (2020)
  082301,
  \href{http://dx.doi.org/10.1103/PhysRevLett.124.082301}{\doi{10.1103/PhysRevLett.124.082301}},
\href{http://www.arXiv.org/abs/1909.01650}{\texttt{arXiv:1909.01650}}.

\bibitem{CMS:2012qk}
\hrefCMSnoop {}{{CMS Collaboration}, ``{Observation of long-range near-side
  angular correlations in proton-lead collisions at the LHC}'',} \textit{ Phys.
  Lett. B} \textbf{ 718} (2013) 795,
  \href{http://dx.doi.org/10.1016/j.physletb.2012.11.025}{\doi{10.1016/j.physletb.2012.11.025}},
\href{http://www.arXiv.org/abs/1210.5482}{\texttt{arXiv:1210.5482}}.

\bibitem{alice:2012qe}
\hrefCMSnoop {}{{ALICE Collaboration}, ``{Long-range angular correlations on
  the near and away side in \pPb\ collisions at $\sqrtsNN = 5.02$\TeV }'',}
  \textit{ Phys. Lett. B} \textbf{ 719} (2013) 29,
  \href{http://dx.doi.org/10.1016/j.physletb.2013.01.012}{\doi{10.1016/j.physletb.2013.01.012}},
\href{http://www.arXiv.org/abs/1212.2001}{\texttt{arXiv:1212.2001}}.

\bibitem{Aad:2012gla}
\hrefCMSnoop {}{{ATLAS Collaboration}, ``Observation of associated near-side
  and away-side long-range correlations in {$\sqrtsNN = 5.02\TeV$} proton-lead
  collisions with the {ATLAS} detector'',} \textit{ Phys. Rev. Lett.} \textbf{
  110} (2013) 182302,
  \href{http://dx.doi.org/10.1103/PhysRevLett.110.182302}{\doi{10.1103/PhysRevLett.110.182302}},
\href{http://www.arXiv.org/abs/1212.5198}{\texttt{arXiv:1212.5198}}.

\bibitem{Aaij:2015qcq}
\hrefCMSnoop {}{{LHCb Collaboration}, ``{Measurements of long-range near-side
  angular correlations in $\sqrtsNN = 5$\TeV proton-lead collisions in the
  forward region}'',} \textit{ Phys. Lett. B} \textbf{ 762} (2016) 473,
  \href{http://dx.doi.org/10.1016/j.physletb.2016.09.064}{\doi{10.1016/j.physletb.2016.09.064}},
\href{http://www.arXiv.org/abs/1512.00439}{\texttt{arXiv:1512.00439}}.

\bibitem{ABELEV:2013wsa}
\hrefCMSnoop {}{{ALICE Collaboration}, ``{Long-range angular correlations of
  pi, K and p in p-Pb collisions at $\sqrtsNN = 5.02\TeV$}'',} \textit{ Phys.
  Lett. B} \textbf{ 726} (2013) 164,
  \href{http://dx.doi.org/10.1016/j.physletb.2013.08.024}{\doi{10.1016/j.physletb.2013.08.024}},
\href{http://www.arXiv.org/abs/1307.3237}{\texttt{arXiv:1307.3237}}.

\bibitem{Khachatryan:2014jra}
\hrefCMSnoop {}{{CMS Collaboration}, ``{Long-range two-particle correlations of
  strange hadrons with charged particles in pPb and PbPb collisions at LHC
  energies}'',} \textit{ Phys. Lett. B} \textbf{ 742} (2015) 200,
  \href{http://dx.doi.org/10.1016/j.physletb.2015.01.034}{\doi{10.1016/j.physletb.2015.01.034}},
\href{http://www.arXiv.org/abs/1409.3392}{\texttt{arXiv:1409.3392}}.

\bibitem{Khachatryan:2015waa}
\hrefCMSnoop {}{{CMS Collaboration}, ``{Evidence for collective multi-particle
  correlations in pPb collisions}'',} \textit{ Phys. Rev. Lett.} \textbf{ 115}
  (2015) 012301,
  \href{http://dx.doi.org/10.1103/PhysRevLett.115.012301}{\doi{10.1103/PhysRevLett.115.012301}},
\href{http://www.arXiv.org/abs/1502.05382}{\texttt{arXiv:1502.05382}}.

\bibitem{Aaboud:2017acw}
\hrefCMSnoop {}{{ATLAS Collaboration}, ``{Measurement of multi-particle
  azimuthal correlations in pp, p+Pb and low-multiplicity Pb+Pb collisions with
  the ATLAS detector}'',} \textit{ Eur. Phys. J. C} \textbf{ 77} (2017) 428,
  \href{http://dx.doi.org/10.1140/epjc/s10052-017-4988-1}{\doi{10.1140/epjc/s10052-017-4988-1}},
\href{http://www.arXiv.org/abs/1705.04176}{\texttt{arXiv:1705.04176}}.

\bibitem{Aaboud:2017blb}
\hrefCMSnoop {}{{ATLAS Collaboration}, ``{Measurement of long-range
  multiparticle azimuthal correlations with the subevent cumulant method in pp
  and p+Pb collisions with the ATLAS detector at the CERN Large Hadron
  Collider}'',} \textit{ Phys. Rev. C} \textbf{ 97} (2018) 024904,
  \href{http://dx.doi.org/10.1103/PhysRevC.97.024904}{\doi{10.1103/PhysRevC.97.024904}},
\href{http://www.arXiv.org/abs/1708.03559}{\texttt{arXiv:1708.03559}}.

\bibitem{Aidala:2016vgl}
\hrefCMSnoop {}{{PHENIX} Collaboration, ``{Measurement of long-range angular
  correlations and azimuthal anisotropies in high-multiplicity p+Au collisions
  at $\sqrtsNN =200\GeV$}'',} \textit{ Phys. Rev. C} \textbf{ 95} (2017)
  034910,
  \href{http://dx.doi.org/10.1103/PhysRevC.95.034910}{\doi{10.1103/PhysRevC.95.034910}},
\href{http://www.arXiv.org/abs/1609.02894}{\texttt{arXiv:1609.02894}}.

\bibitem{PHENIX:2018lia}
\hrefCMSnoop {}{{PHENIX} Collaboration, ``{Creation of quark-gluon plasma
  droplets with three distinct geometries}'',} \textit{ Nature Phys.} \textbf{
  15} (2019) 214,
  \href{http://dx.doi.org/10.1038/s41567-018-0360-0}{\doi{10.1038/s41567-018-0360-0}},
\href{http://www.arXiv.org/abs/1805.02973}{\texttt{arXiv:1805.02973}}.

\bibitem{Adamczyk:2015xjc}
\hrefCMSnoop {}{{STAR} Collaboration, ``{Long-range pseudorapidity dihadron
  correlations in d+Au collisions at $\sqrtsNN = 200$\GeV}'',} \textit{ Phys.
  Lett. B} \textbf{ 747} (2015) 265,
  \href{http://dx.doi.org/10.1016/j.physletb.2015.05.075}{\doi{10.1016/j.physletb.2015.05.075}},
\href{http://www.arXiv.org/abs/1502.07652}{\texttt{arXiv:1502.07652}}.

\bibitem{Adare:2015ctn}
\hrefCMSnoop {}{{PHENIX} Collaboration, ``{Measurements of elliptic and
  triangular flow in high-multiplicity $^{3}$He+Au collisions at $\sqrtsNN =
  200 \GeV$}'',} \textit{ Phys. Rev. Lett.} \textbf{ 115} (2015) 142301,
  \href{http://dx.doi.org/10.1103/PhysRevLett.115.142301}{\doi{10.1103/PhysRevLett.115.142301}},
\href{http://www.arXiv.org/abs/1507.06273}{\texttt{arXiv:1507.06273}}.

\bibitem{Aidala:2017ajz}
\hrefCMSnoop {}{{PHENIX} Collaboration, ``{Measurements of multiparticle
  correlations in d+Au collisions at 200, 62.4, 39, and 19.6 GeV and p+Au
  collisions at 200 GeV and implications for collective behavior}'',} \textit{
  Phys. Rev. Lett.} \textbf{ 120} (2018) 062302,
  \href{http://dx.doi.org/10.1103/PhysRevLett.120.062302}{\doi{10.1103/PhysRevLett.120.062302}},
\href{http://www.arXiv.org/abs/1707.06108}{\texttt{arXiv:1707.06108}}.

\bibitem{Dusling:2015gta}
\hrefCMSnoop {}{K.~Dusling, W.~Li, and B.~Schenke, ``{Novel collective
  phenomena in high-energy proton-proton and proton-nucleus collisions}'',}
  \textit{ Int. J. Mod. Phys. E} \textbf{ 25} (2016) 1630002,
  \href{http://dx.doi.org/10.1142/S0218301316300022}{\doi{10.1142/S0218301316300022}},
\href{http://www.arXiv.org/abs/1509.07939}{\texttt{arXiv:1509.07939}}.

\bibitem{Schlichting:2016sqo}
\hrefCMSnoop {}{{Schlichting, S{\"o}ren and Tribedy, Prithwish},
  ``{Collectivity in small collision systems: An initial-state perspective}'',}
  \textit{ Adv. High Energy Phys.} \textbf{ 2016} (2016) 8460349,
  \href{http://dx.doi.org/10.1155/2016/8460349}{\doi{10.1155/2016/8460349}},
  \href{http://www.arXiv.org/abs/1611.00329}{\texttt{arXiv:1611.00329}}.

\bibitem{Nagle:2018nvi}
\hrefCMSnoop {}{J.~L. Nagle and W.~A. Zajc, ``Small system collectivity in
  relativistic hadronic and nuclear collisions'',} \textit{ Ann. Rev. Nucl.
  Part. Sci.} \textbf{ 68} (2018) 211,
  \href{http://dx.doi.org/10.1146/annurev-nucl-101916-123209}{\doi{10.1146/annurev-nucl-101916-123209}},
\href{http://www.arXiv.org/abs/1801.03477}{\texttt{arXiv:1801.03477}}.

\bibitem{Badea:2019vey}
A.~Badea\hrefCMSnoop {}{ {et~al.}, ``{Measurements of two-particle correlations
  in $\mathrm{e^+e^-}$ collisions at 91 GeV with ALEPH archived data}'',}
  \textit{ Phys. Rev. Lett.} \textbf{ 123} (2019) 212002,
  \href{http://dx.doi.org/10.1103/PhysRevLett.123.212002}{\doi{10.1103/PhysRevLett.123.212002}},
  \href{http://www.arXiv.org/abs/1906.00489}{\texttt{arXiv:1906.00489}}.

\bibitem{ZEUS:2019jya}
\hrefCMSnoop {}{{ZEUS} Collaboration, ``{Two-particle azimuthal correlations as
  a probe of collective behaviour in deep inelastic ep scattering at HERA}'',}
  \textit{ JHEP} \textbf{ 04} (2020) 070,
  \href{http://dx.doi.org/10.1007/JHEP04(2020)070}{\doi{10.1007/JHEP04(2020)070}},
  \href{http://www.arXiv.org/abs/1912.07431}{\texttt{arXiv:1912.07431}}.

\bibitem{Voloshin:1994mz}
\hrefCMSnoop {}{S.~Voloshin and Y.~Zhang, ``{Flow study in relativistic nuclear
  collisions by Fourier expansion of azimuthal particle distributions}'',}
  \textit{ Z. Phys. C} \textbf{ 70} (1996) 665,
  \href{http://dx.doi.org/10.1007/s002880050141}{\doi{10.1007/s002880050141}},
\href{http://www.arXiv.org/abs/hep-ph/9407282}{\texttt{arXiv:hep-ph/9407282}}.

\bibitem{Alver:2010dn}
\hrefCMSnoop {}{B.~H. Alver, C.~Gombeaud, M.~Luzum, and J.-Y. Ollitrault,
  ``{Triangular flow in hydrodynamics and transport theory}'',} \textit{ Phys.
  Rev. C} \textbf{ 82} (2010) 034913,
  \href{http://dx.doi.org/10.1103/PhysRevC.82.034913}{\doi{10.1103/PhysRevC.82.034913}},
\href{http://www.arXiv.org/abs/1007.5469}{\texttt{arXiv:1007.5469}}.

\bibitem{Schenke:2010rr}
\hrefCMSnoop {}{B.~Schenke, S.~Jeon, and C.~Gale, ``Elliptic and triangular
  flow in event-by-event {D=3+1} viscous hydrodynamics'',} \textit{ Phys. Rev.
  Lett.} \textbf{ 106} (2011) 042301,
  \href{http://dx.doi.org/10.1103/PhysRevLett.106.042301}{\doi{10.1103/PhysRevLett.106.042301}},
\href{http://www.arXiv.org/abs/1009.3244}{\texttt{arXiv:1009.3244}}.

\bibitem{Qiu:2011hf}
\hrefCMSnoop {}{Z.~Qiu, C.~Shen, and U.~Heinz, ``{Hydrodynamic elliptic and
  triangular flow in Pb-Pb collisions at $\sqrtsNN = 2.76$\TeV}'',} \textit{
  Phys. Lett. B} \textbf{ 707} (2012) 151,
  \href{http://dx.doi.org/10.1016/j.physletb.2011.12.041}{\doi{10.1016/j.physletb.2011.12.041}},
\href{http://www.arXiv.org/abs/1110.3033}{\texttt{arXiv:1110.3033}}.

\bibitem{Alver:2010gr}
\hrefCMSnoop {}{B.~Alver and G.~Roland, ``{Collision geometry fluctuations and
  triangular flow in heavy-ion collisions}'',} \textit{ Phys. Rev. C} \textbf{
  81} (2010) 054905,
  \href{http://dx.doi.org/10.1103/PhysRevC.81.054905}{\doi{10.1103/PhysRevC.81.054905}},
  \href{http://www.arXiv.org/abs/1003.0194}{\texttt{arXiv:1003.0194}}.
[Erratum: \DOI{10.1103/PhysRevC.82.039903}].

\bibitem{He:2015hfa}
L.~He\hrefCMSnoop {}{ {et~al.}, ``{Anisotropic parton escape is the dominant
  source of azimuthal anisotropy in transport models}'',} \textit{ Phys. Lett.
  B} \textbf{ 753} (2016) 506,
  \href{http://dx.doi.org/10.1016/j.physletb.2015.12.051}{\doi{10.1016/j.physletb.2015.12.051}},
  \href{http://www.arXiv.org/abs/1502.05572}{\texttt{arXiv:1502.05572}}.

\bibitem{Bierlich:2018xfw}
\hrefCMSnoop {}{C.~Bierlich, G.~Gustafson, L.~L{\"o}nnblad, and H.~Shah, ``{The
  Angantyr model for heavy-ion collisions in \PYTHIA8}'',} \textit{ JHEP}
  \textbf{ 10} (2018) 134,
  \href{http://dx.doi.org/10.1007/JHEP10(2018)134}{\doi{10.1007/JHEP10(2018)134}},
  \href{http://www.arXiv.org/abs/1806.10820}{\texttt{arXiv:1806.10820}}.

\bibitem{Kurkela:2019kip}
\hrefCMSnoop {}{A.~Kurkela, U.~A. Wiedemann, and B.~Wu, ``{Flow in AA and pA as
  an interplay of fluid-like and non-fluid like excitations}'',} \textit{ Eur.
  Phys. J. C} \textbf{ 79} (2019) 965,
  \href{http://dx.doi.org/10.1140/epjc/s10052-019-7428-6}{\doi{10.1140/epjc/s10052-019-7428-6}},
  \href{http://www.arXiv.org/abs/1905.05139}{\texttt{arXiv:1905.05139}}.

\bibitem{Andronic:2015wma}
\hrefCMSnoop {}{A.~Andronic {et~al.}, ``{Heavy-flavour and quarkonium
  production in the LHC era: from proton-proton to heavy-ion collisions}'',}
  \textit{ Eur. Phys. J. C} \textbf{ 76} (2016) 107,
  \href{http://dx.doi.org/10.1140/epjc/s10052-015-3819-5}{\doi{10.1140/epjc/s10052-015-3819-5}},
\href{http://www.arXiv.org/abs/1506.03981}{\texttt{arXiv:1506.03981}}.

\bibitem{Dong:2019byy}
\hrefCMSnoop {}{X.~Dong, Y.-J. Lee, and R.~Rapp, ``Open heavy-flavor production
  in heavy-ion collisions'',} \textit{ Ann. Rev. Nucl. Part. Sci.} \textbf{ 69}
  (2019) 417,
  \href{http://dx.doi.org/10.1146/annurev-nucl-101918-023806}{\doi{10.1146/annurev-nucl-101918-023806}},
  \href{http://www.arXiv.org/abs/1903.07709}{\texttt{arXiv:1903.07709}}.

\bibitem{Adare:2006nq}
\hrefCMSnoop {}{{PHENIX} Collaboration, ``{Energy loss and flow of heavy quarks
  in Au+Au collisions at $\sqrtsNN = 200 $\GeV}'',} \textit{ Phys. Rev. Lett.}
  \textbf{ 98} (2007) 172301,
  \href{http://dx.doi.org/10.1103/PhysRevLett.98.172301}{\doi{10.1103/PhysRevLett.98.172301}},
  \href{http://www.arXiv.org/abs/nucl-ex/0611018}{\texttt{arXiv:nucl-ex/0611018}}.

\bibitem{Adamczyk:2017xur}
\hrefCMSnoop {}{{STAR} Collaboration, ``{Measurement of \PDz azimuthal
  anisotropy at midrapidity in Au+Au collisions at $\sqrtsNN = 200\GeV$}'',}
  \textit{ Phys. Rev. Lett.} \textbf{ 118} (2017) 212301,
  \href{http://dx.doi.org/10.1103/PhysRevLett.118.212301}{\doi{10.1103/PhysRevLett.118.212301}},
\href{http://www.arXiv.org/abs/1701.06060}{\texttt{arXiv:1701.06060}}.

\bibitem{Abelev:2014ipa}
\hrefCMSnoop {}{{ALICE Collaboration}, ``{Azimuthal anisotropy of D meson
  production in Pb-Pb collisions at $\sqrtsNN = 2.76\TeV $}'',} \textit{ Phys.
  Rev. C} \textbf{ 90} (2014) 034904,
  \href{http://dx.doi.org/10.1103/PhysRevC.90.034904}{\doi{10.1103/PhysRevC.90.034904}},
\href{http://www.arXiv.org/abs/1405.2001}{\texttt{arXiv:1405.2001}}.

\bibitem{Acharya:2017qps}
\hrefCMSnoop {}{{ALICE Collaboration}, ``{\PD-meson azimuthal anisotropy in
  midcentral Pb-Pb collisions at $\sqrtsNN=5.02$\TeV}'',} \textit{ Phys. Rev.
  Lett.} \textbf{ 120} (2018) 102301,
  \href{http://dx.doi.org/10.1103/PhysRevLett.120.102301}{\doi{10.1103/PhysRevLett.120.102301}},
\href{http://www.arXiv.org/abs/1707.01005}{\texttt{arXiv:1707.01005}}.

\bibitem{Sirunyan:2017plt}
\hrefCMSnoop {}{{CMS Collaboration}, ``{Measurement of prompt \PDz meson
  azimuthal anisotropy in Pb-Pb collisions at $\sqrtsNN = 5.02$\TeV}'',}
  \textit{ Phys. Rev. Lett.} \textbf{ 120} (2018) 202301,
  \href{http://dx.doi.org/10.1103/PhysRevLett.120.202301}{\doi{10.1103/PhysRevLett.120.202301}},
\href{http://www.arXiv.org/abs/1708.03497}{\texttt{arXiv:1708.03497}}.

\bibitem{Khachatryan:2016ypw}
\hrefCMSnoop {}{{CMS Collaboration}, ``{Suppression and azimuthal anisotropy of
  prompt and nonprompt \PJGy production in PbPb collisions at $\sqrtsNN
  =2.76\TeV$}'',} \textit{ Eur. Phys. J. C} \textbf{ 77} (2017) 252,
  \href{http://dx.doi.org/10.1140/epjc/s10052-017-4781-1}{\doi{10.1140/epjc/s10052-017-4781-1}},
\href{http://www.arXiv.org/abs/1610.00613}{\texttt{arXiv:1610.00613}}.

\bibitem{Acharya:2017tgv}
\hrefCMSnoop {}{{ALICE Collaboration}, ``{J/$\psi$ elliptic flow in Pb-Pb
  collisions at $\sqrtsNN=5.02$\TeV}'',} \textit{ Phys. Rev. Lett.} \textbf{
  119} (2017) 242301,
  \href{http://dx.doi.org/10.1103/PhysRevLett.119.242301}{\doi{10.1103/PhysRevLett.119.242301}},
\href{http://www.arXiv.org/abs/1709.05260}{\texttt{arXiv:1709.05260}}.

\bibitem{Sirunyan:2018toe}
\hrefCMSnoop {}{{CMS Collaboration}, ``{Elliptic flow of charm and strange
  hadrons in high-multiplicity pPb collisions at $\sqrtsNN = 8.16\TeV$}'',}
  \textit{ Phys. Rev. Lett.} \textbf{ 121} (2018) 082301,
  \href{http://dx.doi.org/10.1103/PhysRevLett.121.082301}{\doi{10.1103/PhysRevLett.121.082301}},
\href{http://www.arXiv.org/abs/1804.09767}{\texttt{arXiv:1804.09767}}.

\bibitem{Acharya:2017tfn}
\hrefCMSnoop {}{{ALICE Collaboration}, ``{Search for collectivity with
  azimuthal J/$\psi$-hadron correlations in high multiplicity p-Pb collisions
  at $\sqrtsNN = 5.02$ and 8.16\TeV}'',} \textit{ Phys. Lett. B} \textbf{ 780}
  (2018) 7,
  \href{http://dx.doi.org/10.1016/j.physletb.2018.02.039}{\doi{10.1016/j.physletb.2018.02.039}},
\href{http://www.arXiv.org/abs/1709.06807}{\texttt{arXiv:1709.06807}}.

\bibitem{Sirunyan:2018kiz}
\hrefCMSnoop {}{{CMS Collaboration}, ``{Observation of prompt \PJGy meson
  elliptic flow in high-multiplicity pPb collisions at $\sqrtsNN =
  8.16\TeV$}'',} \textit{ Phys. Lett. B} \textbf{ 791} (2019) 172,
  \href{http://dx.doi.org/10.1016/j.physletb.2019.02.018}{\doi{10.1016/j.physletb.2019.02.018}},
\href{http://www.arXiv.org/abs/1810.01473}{\texttt{arXiv:1810.01473}}.

\bibitem{Du:2018wsj}
\hrefCMSnoop {}{X.~Du and R.~Rapp, ``In-medium charmonium production in
  proton-nucleus collisions'',} \textit{ JHEP} \textbf{ 03} (2019) 015,
  \href{http://dx.doi.org/10.1007/JHEP03(2019)015}{\doi{10.1007/JHEP03(2019)015}},
\href{http://www.arXiv.org/abs/1808.10014}{\texttt{arXiv:1808.10014}}.

\bibitem{Zhang:2019dth}
C.~Zhang\hrefCMSnoop {}{ {et~al.}, ``Elliptic flow of heavy quarkonia in {pA}
  collisions'',} \textit{ Phys. Rev. Lett.} \textbf{ 122} (2019) 172302,
  \href{http://dx.doi.org/10.1103/PhysRevLett.122.172302}{\doi{10.1103/PhysRevLett.122.172302}},
\href{http://www.arXiv.org/abs/1901.10320}{\texttt{arXiv:1901.10320}}.

\bibitem{Chatrchyan:2014fea}
\hrefCMSnoop {}{{CMS Collaboration}, ``{Description and performance of track
  and primary-vertex reconstruction with the CMS tracker}'',} \textit{ JINST}
  \textbf{ 9} (2014) P10009,
  \href{http://dx.doi.org/10.1088/1748-0221/9/10/P10009}{\doi{10.1088/1748-0221/9/10/P10009}},
\href{http://www.arXiv.org/abs/1405.6569}{\texttt{arXiv:1405.6569}}.

\bibitem{Chatrchyan:2008zzk}
\hrefCMSnoop {}{{CMS Collaboration}, ``The {CMS} experiment at the {CERN}
  {LHC}'',} \textit{ JINST} \textbf{ 3} (2008) S08004,
\href{http://dx.doi.org/10.1088/1748-0221/3/08/S08004}{\doi{10.1088/1748-0221/3/08/S08004}}.

\bibitem{Khachatryan:2016bia}
\hrefCMSnoop {}{{CMS Collaboration}, ``{The CMS trigger system}'',} \textit{
  JINST} \textbf{ 12} (2017) P01020,
  \href{http://dx.doi.org/10.1088/1748-0221/12/01/P01020}{\doi{10.1088/1748-0221/12/01/P01020}},
  \href{http://www.arXiv.org/abs/1609.02366}{\texttt{arXiv:1609.02366}}.

\bibitem{CMS-PAS-LUM-17-002}
\href {http://cds.cern.ch/record/2628652}{{CMS Collaboration}, ``{CMS
  luminosity measurement using 2016 proton-nucleus collisions at
  nucleon-nucleon center-of-mass energy of 8.16 TeV}'',} Physics Analysis
  Summary CMS-PAS-LUM-17-002, CERN, Geneva, 2018.

\bibitem{Sirunyan:2017quh}
\hrefCMSnoop {}{{CMS Collaboration}, ``Constraints on the chiral magnetic
  effect using charge-dependent azimuthal correlations in $\mathrm{pPb}$ and
  pbpb collisions at the {CERN Large Hadron Collider}'',} \textit{ Phys. Rev.
  C} \textbf{ 97} (2018) 044912,
  \href{http://dx.doi.org/10.1103/PhysRevC.97.044912}{\doi{10.1103/PhysRevC.97.044912}},
\href{http://www.arXiv.org/abs/1708.01602}{\texttt{arXiv:1708.01602}}.

\bibitem{Sirunyan:2017uyl}
\hrefCMSnoop {}{{CMS Collaboration}, ``{Observation of correlated azimuthal
  anisotropy fourier harmonics in pp and p+Pb collisions at the LHC}'',}
  \textit{ Phys. Rev. Lett.} \textbf{ 120} (2018) 092301,
  \href{http://dx.doi.org/10.1103/PhysRevLett.120.092301}{\doi{10.1103/PhysRevLett.120.092301}},
\href{http://www.arXiv.org/abs/1709.09189}{\texttt{arXiv:1709.09189}}.

\bibitem{Zyla:2020zbs}
\hrefCMSnoop {}{{Particle Data Group} Collaboration, ``{Review of Particle
  Physics}'',} \textit{ PTEP} \textbf{ 2020} (2020), no.~8, 083C01,
  \href{http://dx.doi.org/10.1093/ptep/ptaa104}{\doi{10.1093/ptep/ptaa104}}.

\bibitem{Hocker:2007ht}
\href {http://pos.sissa.it/archive/conferences/050/040/ACAT_040.pdf}{H.~Voss,
  A.~H{\"o}cker, J.~Stelzer, and F.~Tegenfeldt, ``{TMVA} --- the toolkit for
  multivariate data analysis'',} in \textit{ XIth International Workshop on
  Advanced Computing and Analysis Techniques in Physics Research (ACAT)},
  p.~40.
\newblock 2009.
\newblock
\href{http://www.arXiv.org/abs/physics/0703039}{\texttt{arXiv:physics/0703039}}.
\newblock

\bibitem{Sjostrand:2014zea}
T.~Sj{\"o}strand\hrefCMSnoop {}{ {et~al.}, ``{An introduction to \PYTHIA
  8.2}'',} \textit{ Comput. Phys. Commun.} \textbf{ 191} (2015) 159,
  \href{http://dx.doi.org/10.1016/j.cpc.2015.01.024}{\doi{10.1016/j.cpc.2015.01.024}},
\href{http://www.arXiv.org/abs/1410.3012}{\texttt{arXiv:1410.3012}}.

\bibitem{Khachatryan:2015pea}
\hrefCMSnoop {}{{CMS Collaboration}, ``Event generator tunes obtained from
  underlying event and multiparton scattering measurements'',} \textit{ Eur.
  Phys. J. C} \textbf{ 76} (2016) 155,
  \href{http://dx.doi.org/10.1140/epjc/s10052-016-3988-x}{\doi{10.1140/epjc/s10052-016-3988-x}},
\href{http://www.arXiv.org/abs/1512.00815}{\texttt{arXiv:1512.00815}}.

\bibitem{Pierog:2013ria}
T.~Pierog\hrefCMSnoop {}{ {et~al.}, ``{EPOS} {LHC}: {Test} of collective
  hadronization with data measured at the {CERN} {Large Hadron Collider}'',}
  \textit{ Phys. Rev. C} \textbf{ 92} (2015) 034906,
  \href{http://dx.doi.org/10.1103/PhysRevC.92.034906}{\doi{10.1103/PhysRevC.92.034906}},
\href{http://www.arXiv.org/abs/1306.0121}{\texttt{arXiv:1306.0121}}.

\bibitem{Chatrchyan:2013nka}
\hrefCMSnoop {}{{CMS} Collaboration, ``{Multiplicity and transverse momentum
  dependence of two- and four-particle correlations in \pPb\ and \PbPb\
  collisions}'',} \textit{ Phys. Lett. B} \textbf{ 724} (2013) 213,
  \href{http://dx.doi.org/10.1016/j.physletb.2013.06.028}{\doi{10.1016/j.physletb.2013.06.028}},
\href{http://www.arXiv.org/abs/1305.0609}{\texttt{arXiv:1305.0609}}.

\bibitem{CrystalBallRef}
\href {http://www.slac.stanford.edu/cgi-wrap/getdoc/slac-r-236.pdf}{M.~J.
  Oreglia, ``A study of the reactions $\psi^\prime \to \gamma \gamma \psi$''}.
\newblock PhD thesis, Stanford University, 1980.
\newblock {SLAC} Report {SLAC-R-236}.

\bibitem{Nahrgang:2014vza}
M.~Nahrgang\hrefCMSnoop {}{ {et~al.}, ``Elliptic and triangular flow of heavy
  flavor in heavy-ion collisions'',} \textit{ Phys. Rev. C} \textbf{ 91} (2015)
  014904,
  \href{http://dx.doi.org/10.1103/PhysRevC.91.014904}{\doi{10.1103/PhysRevC.91.014904}},
\href{http://www.arXiv.org/abs/1410.5396}{\texttt{arXiv:1410.5396}}.

\bibitem{He:2014cla}
\hrefCMSnoop {}{M.~He, R.~J. Fries, and R.~Rapp, ``Heavy flavor at the {Large
  Hadron Collider} in a strong coupling approach'',} \textit{ Phys. Lett. B}
  \textbf{ 735} (2014) 445,
  \href{http://dx.doi.org/10.1016/j.physletb.2014.05.050}{\doi{10.1016/j.physletb.2014.05.050}},
\href{http://www.arXiv.org/abs/1401.3817}{\texttt{arXiv:1401.3817}}.

\bibitem{Xu:2015bbz}
\hrefCMSnoop {}{J.~Xu, J.~Liao, and M.~Gyulassy, ``Bridging soft-hard transport
  properties of quark-gluon plasmas with {CUJET3.0}'',} \textit{ JHEP} \textbf{
  02} (2016) 169,
  \href{http://dx.doi.org/10.1007/JHEP02(2016)169}{\doi{10.1007/JHEP02(2016)169}},
\href{http://www.arXiv.org/abs/1508.00552}{\texttt{arXiv:1508.00552}}.

\bibitem{Zhang:2020ayy}
C.~Zhang\hrefCMSnoop {}{ {et~al.}, ``{Collectivity of heavy mesons in
  proton-nucleus collisions}'',} \textit{ Phys. Rev. D} \textbf{ 102} (2020)
  034010,
  \href{http://dx.doi.org/10.1103/PhysRevD.102.034010}{\doi{10.1103/PhysRevD.102.034010}},
  \href{http://www.arXiv.org/abs/2002.09878}{\texttt{arXiv:2002.09878}}.

\bibitem{Lange:2001uf}
\hrefCMSnoop {}{D.~J. Lange, ``{The \EVTGEN particle decay simulation
  package}'',} \textit{ Nucl. Instrum. Meth. A} \textbf{ 462} (2001) 152,
\href{http://dx.doi.org/10.1016/S0168-9002(01)00089-4}{\doi{10.1016/S0168-9002(01)00089-4}}.

\bibitem{Aad:2020grf}
\hrefCMSnoop {}{{ATLAS Collaboration}, ``{Measurement of azimuthal anisotropy
  of muons from charm and bottom hadrons in Pb+Pb collisions at $\sqrt
  {s_{NN}}$ = 5.02 TeV with the ATLAS detector}'',} \textit{ Phys. Lett. B}
  \textbf{ 807} (2020) 135595,
  \href{http://dx.doi.org/10.1016/j.physletb.2020.135595}{\doi{10.1016/j.physletb.2020.135595}},
  \href{http://www.arXiv.org/abs/2003.03565}{\texttt{arXiv:2003.03565}}.

\end{thebibliography}\endgroup
\cleardoublepage \appendix\section{The CMS Collaboration \label{app:collab}}\begin{sloppypar}\hyphenpenalty=5000\widowpenalty=500\clubpenalty=5000\vskip\cmsinstskip
\textbf{Yerevan Physics Institute, Yerevan, Armenia}\\*[0pt]
A.M.~Sirunyan$^{\textrm{\dag}}$, A.~Tumasyan
\vskip\cmsinstskip
\textbf{Institut f\"{u}r Hochenergiephysik, Wien, Austria}\\*[0pt]
W.~Adam, F.~Ambrogi, T.~Bergauer, M.~Dragicevic, J.~Er\"{o}, A.~Escalante~Del~Valle, M.~Flechl, R.~Fr\"{u}hwirth\cmsAuthorMark{1}, M.~Jeitler\cmsAuthorMark{1}, N.~Krammer, I.~Kr\"{a}tschmer, D.~Liko, T.~Madlener, I.~Mikulec, N.~Rad, J.~Schieck\cmsAuthorMark{1}, R.~Sch\"{o}fbeck, M.~Spanring, W.~Waltenberger, C.-E.~Wulz\cmsAuthorMark{1}, M.~Zarucki
% ...
\vskip\cmsinstskip
\dag: Deceased\\
1:  Also at Vienna University of Technology, Vienna, Austria\\
% ...
\end{sloppypar}
\end{document}